\documentclass[a4paper]{article}
\usepackage[utf8x]{inputenc}
\usepackage[english]{babel}
\usepackage{slashed}
\usepackage[T1]{fontenc}
\usepackage{cite}
\usepackage{float}
\usepackage[font=small,labelfont=md]{caption}
\usepackage{hyperref}
\usepackage{epstopdf}
\usepackage{amsmath,amsthm,amsfonts,amssymb,latexsym}
 \usepackage[english]{babel}
\usepackage[dvips]{graphicx}
\usepackage{amsfonts}
\usepackage{textcomp}
\usepackage{subfig}
\usepackage{pstricks}
\usepackage{authblk}
\usepackage[normalem]{ulem}
\usepackage{color}
\definecolor{orange}{rgb}{1,0.5,0}

\title{Top quark anomalous tensor couplings in the two-Higgs-doublet models}

\author[1]{Luc\'ia Duarte\thanks{lucia@fisica.edu.uy}}
\author[2]{Gabriel A. Gonz\'alez-Sprinberg\thanks{gabrielg@fisica.edu.uy}}
\author[3]{Jordi Vidal\thanks{vidal@uv.es}}

\affil[1]{Instituto de F\'\i sica, Facultad de Ingenier\'\i a, Universidad de la
Rep\'ublica, Uruguay.}
\affil[2]{Instituto de F\'\i sica, Facultad de Ciencias, Universidad de la
Rep\'ublica,  Uruguay.}
\affil[3]{Departament de F\'\i sica Te\`orica Universitat de Val\`encia, E-46100
Burjassot, Val\`encia, Spain, and 
Instituto de F\'\i sica Corpuscular (IFIC), Centro Mixto Universitat de Val\`encia-CSIC, Val\`encia, Spain.}

\begin{document}

\maketitle
\vspace {-13cm}
\hfill FTUV-13-16048
\vspace {11,8cm}

\begin{abstract}

We compute the one loop right and left anomalous tensor couplings ($g_R$ and $g_L$, respectively) for the top quark, in the aligned two-Higgs-doublet model. They are the magnetic-like couplings  in the most general parameterization of the 
$tbW$ vertex. We find that the aligned two-Higgs doublet model, that includes as particular cases some of the most studied extensions of the Higgs sector,  introduces new electroweak contributions and  provides theoretical predictions that are very sensitive to both new scalar 
masses and the neutral scalar mixing angle. For a large area in the parameters space
 we obtain significant deviations in both the real and the imaginary parts of the couplings $g_R$ and $g_L$, compared to the predictions given by the electroweak sector of the Standard Model.
The most important ones are those involving the imaginary part of the left coupling $g_{L}$ and the real part of the right coupling $g_R$. The real part of $g_L$ and the imaginary part of $g_R$ also show an important sensitivity to new physics scenarios. The model can also account for new CP violation effects via the introduction 
of complex alignment parameters that have important consequences on the values for the imaginary parts of the couplings. 
The top anomalous tensor couplings will be measured at the LHC and at future colliders providing a complementary insight on new physics, independent from the bounds in top decays coming from B physics and $b \rightarrow s \gamma$.

\end{abstract}

\section{Introduction}

The recent discovery at the LHC of a new neutral boson \cite{:2012gk,:2012gu} points to a spontaneous electroweak  symmetry 
breaking mechanism involving scalars, but more experimental analyses are still needed to distinguish whether we face the unique 
Standard Model (SM) Higgs boson or an extended scalar sector. Besides, top quark physics at the LHC can play a role in this quest, as it is 
expected to probe physics 
beyond the electroweak scale. 

While no deviation from the SM predictions has been found in top physics yet \cite{Abazov:2010jn,Aad:2012ky,Chatrchyan:2013jna,ATLAS-CONF-2013-032}, the angular distribution in the dominant decay mode 
$t \rightarrow b W^{+}$ is going to be precisely measured at the LHC. This measurement can probe
the SM beyond tree-level and might be sensitive to new physics in the electroweak sector, where it is expected to appear.
Besides, new physics interactions might show up in the measurements of the top anomalous couplings because 
they may modify the strength and structure of the $tbW$ vertex. The SM one loop predictions for the anomalous tensor couplings receive contributions from QCD, coming from gluon
exchange, and from the  electroweak (EW) sector of the SM. The real parts of the couplings receive contributions from  both QCD and the EW sectors, while the imaginary parts are
generated exclusively by the EW corrections.
The electroweak  contributions to the 
right and left top anomalous tensor couplings, $g_R$ and $g_L$ respectively, have recently been calculated in Ref. \cite{GonzalezSprinberg:2011kx, GonzalezSprinberg:2013er}. These quantum corrections amount to 
 19\% of the dominant QCD contribution for the real part of the right coupling $ g_{R}$, and 9\% for the 
real part of the left coupling $ g_{L}$. This last
 prediction is in slight tension with existing indirect constraints obtained recently from $B$ decays data
 \cite{Drobnak:2011aa,Grzadkowski:2008mf}. Besides, in Ref. \cite{GonzalezSprinberg:2011kx, GonzalezSprinberg:2013er} the imaginary parts of $g_R$ and $g_L$ and the real part of $g_L$ were also calculated. Direct constraints on the top anomalous couplings were obtained by D0 \cite{Abazov:2010jn} at
 Tevatron, and by ATLAS and CMS \cite{Aad:2012ky,ATLAS-CONF-2013-032,Chatrchyan:2013jna} at the LHC. These last are
 still looser than the indirect ones, but a much better sensitivity is expected in the LHC measurements in the future \cite{AguilarSaavedra:2010nx}. 
 
Among the SM extensions, the inclusion of one extra scalar doublet is a minimal choice and results in a variety of dynamical possibilities. Two-Higgs-doublet models (2HDM) can also be read as a low energy effective theory. Besides, they provide new ways to introduce CP violation sources, both in
 the scalar potential and in the Yukawa sector. Many new physics scenarios, including supersymmetry, can lead to a low energy spectrum containing
 the SM fields, plus additional scalar multiplets. In general, 2HDMs allow the appearance of unwanted flavor changing neutral currents (FCNCs) 
 unless ad-hoc restrictions are imposed at the lagrangian, such as $\mathcal{Z}_{2}$ symmetries 
 that, in addition, also forbid CP violation in the scalar potential. These so-called natural flavor conservation models include the well known 
types I, II, III, X, Y and the inert 2HDM. For a comprehensive review see \cite{Branco:2011iw}. A less restrictive and 
more general alternative is given by the aligned two-Higgs-doublet
model (A2HDM) \cite{Pich:2009sp}, which imposes the proportionality of both Yukawa matrices, with complex alignment parameters, and includes all the previously mentioned models as particular limits. These complex alignment parameters allow for a new CP violation source in the Yukawa sector, independently of the form of the potential, and in addition to the Cabibbo-Kobayashi-Maskawa quark mixing matrix. 

Previous works have studied the top quark decay vertex in the context of the 2HDM (type II), the MSSM, little Higgs and TC2 models 
\cite{Beneke:2000hk,Tait:2000sh,Bernreuther:2008us}.
In this paper we calculate the  top quark anomalous tensor couplings in the general framework of the A2HDM. Our work 
complements the flavor physics analysis where this model has been thoroughly studied \cite{Jung:2010ik,Jung:2010ab,Jung:2012vu,Celis:2012dk}, 
and the recent work \cite{Celis:2013rcs} that takes into account the LHC measurements.

In section \ref{A2HDMov} we briefly review the A2HDM and comment on the constraints for its parameters. In section \ref{gesA2HDM} we define the
vertex parameterization and, in section \ref{Observables}, we review the theoretical and experimental status of the top anomalous tensor couplings. Our analytical calculation is
 introduced in section \ref{OurCalc} and numerical results for the different scalars mass scenarios chosen is presented in section \ref{Res}. In particular, we compare the A2HDM predictions for the top anomalous tensor  couplings to the recently calculated 
electroweak values. 
We investigate the sensitivity of the anomalous tensor 
couplings ($g_R$ in section \ref{cgr} and $g_L$ in section \ref{cgl}) to the scalars mixing angle and alignment parameters, 
taking a CP conserving scalar potential, but allowing the presence of complex 
CP violating phases. The results for the Type I and II 2HDM are also shown in section \ref{IyII}. Finally, we present our 
conclusions in section \ref{Conc}. 

\section{A2HDM: overview}\label{A2HDMov}

The 2HDMs extend the SM by adding a second scalar doublet of hypercharge Y=1/2. The EW sector in these models is significantly different from the SM. 
The A2HDM model incorporates, in addition to three Goldstone bosons, five physical scalars: two charged scalar fields $H^{\pm}(x)$ and three neutral scalars 
$ \{ \varphi_{i}(x) \}_{i=1,2,3} =\{h(x),H(x),A(x)\}$, related through an orthogonal transformation $\mathcal{R}$ to the gauge eigenstates $S_{i}$: 

\begin{equation}
\varphi_{i}(x)= \mathcal{R}_{ij}S_{j}(x) \; ; \,\;\, i, j = 1, 2, 3.
\end{equation}
 The mixing matrix $\mathcal{R}$ depends on the particular form of the potential, which is also responsible of the structure of the scalars mass matrix and mass eigenstates. Taking a CP conserving potential and in the so-called Higgs basis, where only one doublet acquires a nonzero vacuum expectation value, the mixing matrix is written as:

\begin{equation}\label{matrix}
 \left( \begin{array}{c}
H \\
h \\
A \end{array} \right)= \left( \begin{array}{ccc}
cos\gamma & sin\gamma & 0\\
-sin\gamma & cos\gamma & 0\\
0 & 0 & 1
 \end{array} \right) 
 \left( \begin{array}{c}
S_{1} \\
S_{2}\\
S_{3} \end{array} \right), 
\end{equation}
where $\gamma$ is the neutral scalars mixing angle.

The generic Yukawa lagrangian, with standard fermionic content, gives rise to FCNCs because the Yukawa couplings of both doublets cannot be simultaneously diagonalized. 
Tree-level FCNCs can be avoided by requiring the \emph{alignment} in flavor space of the Yukawa couplings, {\it i.e.}, by making both Yukawa matrices to be proportional to each other, for each type of right handed fermion. If, in addition, the proportionality parameters $\varsigma_{f}$ ($f\equiv u,d,l$) are taken to be arbitrary complex numbers then new sources of CP violation are introduced.

In the mass eigenstates basis the Yukawa lagrangian is written as:
\begin{align}\label{LYukawaA2HDM}
 \mathcal{L}_{Y}= &-\frac{\sqrt{2}}{v} H^{+}(x)\bar u(x)[\varsigma_{d}VM_{d}P_{R}-\varsigma_{u}M_{u}VP_{L}]d(x)\nonumber \\
 \qquad {} &
 -\frac{\sqrt{2}}{v}H^{+}(x)\varsigma_{l}\bar \nu(x) M_{l}P_{R}l(x)\nonumber \\
 \qquad {} &
 -\frac{1}{v}\sum_{i,f}\varphi_{i}(x)y^{\varphi_{i}}_{f}\bar f(x)M_{f}P_R f(x)+ h.c.\ ,
\end{align}
where $V$ is the Cabibbo-Kobayashi-Maskawa matrix, $P_{R,L}\equiv\frac{1}{2}(1\pm\gamma_{5})$ are the chirality projectors and $M_f$ ($f\equiv u,d,l$) are the diagonal mass matrices.

The neutral Yukawa terms are flavor-diagonal and the couplings $y^{\varphi_{i}}_{f}$ are proportional to the corresponding elements of the neutral scalar mixing matrix $\mathcal{R}$:
\begin{equation}
y^{\varphi_{i}}_{d,l}= \mathcal{R}_{i1}+(\mathcal{R}_{i2}+i\mathcal{R}_{i3})\varsigma_{d,l}\ , \qquad y^{\varphi_{i}}_{u}= \mathcal{R}_{i1}+(\mathcal{R}_{i2}-i\mathcal{R}_{i3})\varsigma^*_{u}\ .
\end{equation}

The A2HDM leads to a structure where all fermion-scalar interactions are proportional to the fermion masses, giving rise to a hierarchy of non tree-level FCNC effects. 
 Bounds on the $\varsigma_{f}$ parameters have been explored in Ref. \cite{Jung:2010ik}, in the context of charged Higgs phenomenology. There, constraints on the $\varsigma_{u,d,l}$ parameters as a function of $m_{H^{\pm}}$ were obtained from lepton decays and leptonic and semi-leptonic tree-level decays of pseudoscalar mesons. From tau decays they obtained that $|\varsigma_{l}|/m_{H^{\pm}} \leq 0.40$ GeV$^{-1}$ and from a global fit to leptonic and semi-leptonic decays they got the bounds $|\varsigma_{u} \varsigma^{*}_{l}|/m^{2}_{H^{\pm}} \lesssim 0.01$ GeV$^{-2}$ and $|\varsigma_{d} \varsigma^*_{l}|/m^{2}_{H^{\pm}}< 0.1$ GeV$^{-2}$

Bounds on $|\varsigma_{u}|$, obtained from the top quark loops contributions in $Z\rightarrow b \bar b$ decays, give $|\varsigma_{u}|< 0.91(1.91)$ for $m_{H^{\pm}}=80 (500)$ GeV \cite{Jung:2010ik}. Mixing processes, such as $B^{0}-\bar B^{0}$ and $K^{0}-\bar K^{0}$ mixing, result in less restrictive limits because they depend on the relative phase between the alignment parameters $\varsigma_{u}$ and $\varsigma_{d}$. From the radiative decay $\bar B\rightarrow X_{s}\gamma$, the bound $|\varsigma_{u}||\varsigma_{d}|\lesssim 20$, for $M_{H^\pm}\in(80,500)$ GeV, is obtained by assuming $|\varsigma_{u}|<3$ \cite{Jung:2010ik}. More recently, the observed excess in $\tau$ lepton production in semileptonic B-meson decays reported by BaBar \cite{Lees:2012xj} has been analyzed within the A2HDM context and it points towards bigger values for the product $|\varsigma^*_{l}\varsigma_{u}|/m_{H^{\pm}}$ than the one previously obtained in Ref. \cite{Celis:2012dk}. 
A recent paper \cite{Celis:2013rcs} analyzes the A2HDM in the light of the "Higgs-like" particle discovery \cite{:2012gk,:2012gu} and Higgs signals data from Tevatron \cite{Aaltonen:2012qt} getting some constraints even with the large experimental uncertainties existing up to now. The authors search for possible ways to enhance the diphoton channel while being compatible with the rest of the data. Although a pure CP-odd assignment for the new particle is ruled out, they investigate several possibilities including the CP conserving $\mathcal{Z}_{2}$ limit, degenerate CP violating mixtures in the scalar potential and charged scalars contributions to the amplitude $h \rightarrow 2 \gamma$. Concerning the A2HDM an enhancement is obtained in the $\gamma \gamma$ rate with a complex top Yukawa coupling with real part close to the SM value.

\section{Top tensor couplings in the A2HDM}\label{gesA2HDM}

The $tbW^{+}$ vertex can be studied by parameterizing the amplitude $\mathcal{M}_{tbW}$ of the $t(p)\rightarrow b(p') W^{+}(q)$ decay with the most general Lorentz structure, for on-shell particles, in the following way:
\begin{eqnarray}
\mathcal{M}_{tbW}&=& - \frac{e}{\sin\theta_{w}\sqrt{2}}\epsilon^{\mu*} \, \times \nonumber \\
&& \overline{u}_{b}(p')\left[\gamma_{\mu}(V_{L} P_{L}+V_{R} P_{R})+ 
\frac{i \sigma_{\mu\nu}q^{\nu}}{M_{W}}( g_{L} P_{L}+ g_{R} P_{R})\right]u_{t}(p),\label{gesdef}
\end{eqnarray}
where the outgoing $W^{+}$ momentum, mass and polarization vector are $q=p-p'$, $M_{W}$ and $\epsilon^{\mu*}$, respectively.
The form factors are all dimensionless; $V_{L}$ and $V_{R}$ parameterize the vector and axial-vector couplings while $ g_{L}$ and $ g_{R}$ are the so called left and right anomalous tensor couplings, respectively.

The expression (\ref{gesdef}) is the most general model independent parameterization for the $tbW^{+}$ vertex. Another approach to the problem is the effective lagrangian method. This technique describes the low energy physics of a theory by using non-renormalized terms in the lagrangian written with the SM fields and invariant under the gauge symmetry of the SM. This approach assumes that the new physics spectrum is very well above the EW energy scale \cite{Buchmuller:1985jz, AguilarSaavedra:2008zc, Kane:1991bg}. In this paper we will adopt the first approach that does not rely on any particular assumptions and that it is compatible with the energy scales explored by the LHC.

The tree level SM values for the couplings are $V_L=V_{tb}$ and
$V_R = g_R = g_L = 0$. All these couplings receive corrections at one loop in
the SM and in extended models. The measurement of $V_L=V_{tb} \simeq 1$ is
still affected by large uncertainties \cite{Beringer:1900zz} and its
determination may be an open window to test new physics, but this issue
and any possible deviations of $V_R$ from 0 will not be the target of
this work, where we concentrate only on the tensor couplings $g_R$
and $g_L$.

 Within the SM, the dominant contribution to the real part of the 
 tensor couplings comes from the QCD one loop diagram, generated by gluon exchange. 
The computed values for $g^{QCD}_R$ \cite{Li:1990qf} and for $g^{QCD}_{L}$ \cite{GonzalezSprinberg:2011kx, GonzalezSprinberg:2013er}, for $m_{t}=171$ GeV, are:
\begin{equation}
g_{R}^{QCD} = - 6.61 \times 10^{-3}, \qquad g_{L}^{QCD} = - 1.118 \times 10^{-3}.
\label{qcd-contr}
\end{equation}  
Both the real and the imaginary one loop corrections in the EW sector  
for the SM were obtained in Ref. \cite{GonzalezSprinberg:2011kx, GonzalezSprinberg:2013er} for a SM Higgs $h^0$ with $m_{h^0}=150$ GeV. The  values of the electroweak contribution to these couplings in the SM but for the now measured value $m_{h^{0}}=126 $ GeV are:
\begin{equation}
g_{R}^{EW} = - (1.24+1.23i) \times 10^{-3}, \qquad g_{L}^{EW} = - (0.102+0.014i) \times 10^{-3}.
\label{ew-contr}
\end{equation} 
The imaginary numbers come exclusively from absorptive parts in some of the EW diagrams. 
The complete SM one loop contributions, including the one loop QCD and the electroweak SM contributions for the tensor couplings are then:
\begin{equation}
 g^{SM}_{R} = - (7.85+1.23i) \times 10^{-3}, \qquad g^{SM}_{L} = - (1.220+0.014i) \times 10^{-3}.
\label{sm-contr}
\end{equation}
 
From Eqs. (\ref{ew-contr}) and (\ref{sm-contr}) it can be seen that for the real part of $g_{R}$ and $g_L$, the EW contribution is 16\% and 8\%,
respectively, of the total SM values. To be sensitive to new physics in the real part of the couplings, one has to be accurate  enough  in the measurement to
disentangle the QCD and EW contributions. Instead,  the imaginary parts are directly sensitive to new physics, because they come only from
the EW diagrams.  Note that the imaginary right coupling is of the same order of magnitude than the real parts of both couplings;  however, for the left
coupling, the imaginary part is very small.
Because the anomalous tensor couplings are chirality flipping quantities, the EW contributions to $g_{L}$ are lower in magnitude than those for $g_{R}$, due to the flow of chirality in the diagrams with the standard $tbW$ vertex. 

\subsection{Observables and experimental status}\label{Observables}

Besides the observables that were already considered in the literature -branching ratios, helicity fractions, angular distributions and asymmetries \cite{Lampe:1995xb, delAguila:2002nf}- new observables have been defined in Ref. \cite{AguilarSaavedra:2010nx}. These make use of the spin properties of the polarized top quarks produced at the LHC in order to define quantities that are sensitive to the imaginary parts of the anomalous tensor couplings. The imaginary part of the anomalous tensor couplings is not a CP-odd quantity, but CP violation can nevertheless be investigated by comparing the properties of the top and anti-top decay vertex: a change in the sign of the imaginary parts of the decay tensor couplings ($Im(g_R)$ and/or $Im (g_L)$), when comparing top and anti-top, would point out to CP violation.
 
The normal asymmetry $A^{N}_{FB}$, defined in \cite{AguilarSaavedra:2010nx}, considers the orthogonal direction to the plane defined by the top spin and the W momenta. The forward and backward directions are defined with respect to the angle of the charged lepton (into which the W decays) momenta in the W rest frame with the W momenta in the top rest frame. This asymmetry vanishes for real anomalous couplings and, consequently, it turns out to be very sensitive to $Im ( g_{R})$. For small $ g_{R}$ and taking $V_{L}=1$, $V_{R}= g_{L}=0$, the authors obtain $A^{N}_{FB}=0.64 \cdot P \cdot Im ( g_{R})$, for top quarks with normal polarization degree $P$. A combined analysis of this new observable together with the usual $W$ helicity fractions, the asymmetries in the top quark rest frame, and the $tW$ cross section, allows a model independent fit of all the $tbW$ vertex parameters \cite{AguilarSaavedra:2010nx,Rindani:2011pk}. Preliminary measurements of the normal forward-backward asymmetry using data up to 
2011 at the LHC give the bound $-0.07\leq Im ( g_{R})\leq 0.18$ at 68\% CL \cite{ATLAS-CONF-2013-032}.

ATLAS and CMS have recently published bounds for the top anomalous couplings. They analyzed data obtained in 2010 and 2011 on the $W$ helicity fractions in top pair events. The CMS bounds \cite{Chatrchyan:2013jna} are given in two different scenarios: (i) assuming $V_{L}=1,V_{R}= g_{L}=0$ and leaving $Re( g_{R})$ as a free parameter, and (ii) leaving $V_{R}$ to be free. The first assumption (i) gives the best fitted value: $Re(g_{R})=-0.008\pm 0.024(stat.)^{+0.029}_{-0.030}(syst.)$.
ATLAS limits $-0.14<Re( g_{L})<0.11$ and $-0.08<Re( g_{R})<0.04$, at 95\% CL\cite{Aad:2012ky}, were obtained assuming all anomalous couplings set to zero, except the one to be bounded.

Direct bounds for the top anomalous couplings are also available from Tevatron. D0 Collaboration\cite{:2012iwa} performed a combined analysis of the measurements of the $W$ bosons helicities \cite{Abazov:2010jn} and those of the single top quark production \cite{Abazov:2011pm}. Taking real anomalous form factors, they studied the allowed regions of the squared modulus of a form factor, $| g_{L,R}|^2$ or, alternatively, $|V_{R}|^2$, as a function of $|V_{L}|^2$ with all other couplings set to zero. 
 A different analysis of early LHC data can be found in Ref. \cite{AguilarSaavedra:2011ct} where they combined recent measurements in ATLAS of top quark decay asymmetries with the t-channel single top cross section measured by CMS. This combination of data allows a better determination of the anomalous couplings bounds: they plotted the allowed regions in the $ (g_{L} , g_{R})$ plane at 95\% CL, assuming $V_{L}=1,V_{R}=0$ and that both tensor couplings are real, resulting in the limits $|g_L|\leq 0.45$, and $-0.55\leq g_R\leq 0.20$ or $0.70\leq g_R\leq 0.90$.

Finally, indirect bounds were obtained \cite{Grzadkowski:2008mf, Drobnak:2011aa} from the $Br(B \rightarrow X_{s}\gamma)$ branching ratio, measured at B factories. These last results use the most recent global fits in neutral mesons oscillations \cite{Ligeti:2010ia, Lenz:2010gu} and represent the strongest bounds on the anomalous tensor couplings. In our notation, they get the following bounds: $-0.001<Re( g_{L})<0.0003$ and $-0.07<Re( g_{R})<0.27$, at 95\% CL, assuming real couplings, $V_{L}=1$ and only one non-vanishing form factor at a time.

\subsection{Our calculation}\label{OurCalc}

Let us start discussing the general features of the $t\rightarrow bW^{+}$ decay in the A2HDM. We will concentrate in the EW part because it is the only one that is different from the SM. 

At tree-level the fermion couplings to gauge bosons are not modified with respect to the SM by the presence of a new scalar doublet, so that the $tbW$ vertex remains unchanged. However, at one loop, besides the SM fields (top and bottom quarks, gluons, gauge bosons $W$, $Z$ and $\gamma$, and Goldstone bosons $G^{0 \pm}$), we have to consider the contributions of the new fields (the three neutral scalars $h$, $H$ and $A$, and the charged scalars $H^{\pm}$) circulating in the internal lines of the loop.

\begin{figure}[ht]
\begin{center}
\includegraphics[width=0.9\textwidth]{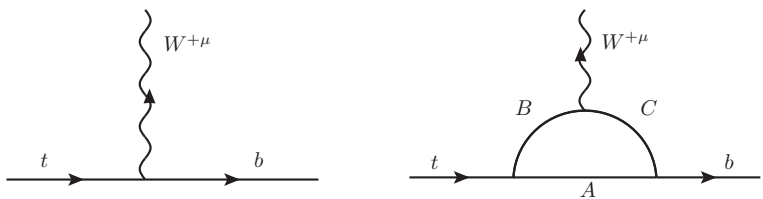}
\caption{The $t\rightarrow bW^{+}$ vertex: tree-level and one loop diagrams.}\label{tbwgenerico}
\end{center}
\end{figure}

There is only one topology of the one 
loop diagrams that contributes to the vertex and, thus, to the anomalous tensor couplings. This is shown in Fig.\ref{tbwgenerico}. 

The different Feynman diagrams in the calculation are identified by naming the particles circulating in the loop as ABC.
We show in Table \ref{tabla} the diagrams classified by the position taken by the neutral scalars $\varphi_{i}$: in type (a) the neutral scalars take position A , in (b), (d) and (f) they are in position B, with a t quark in the loop, and in (c), (e) and (g) they are in position C, with a b quark in the loop. Depending on the value for the mass of the charged scalar $H^+$, diagrams type (c) can develop absorptive parts. In addition, diagrams type (e) and (g) always have an imaginary part.

We recover the SM values from the A2HDM just taking the $m_{H, A, H^{+}}\rightarrow \infty$ limit and identifying $h\equiv h^{0}$. In that limit we explicitly checked that the contributions to the anomalous tensor form factors in the A2HDM are identical to the EW corresponding ones obtained in \cite{GonzalezSprinberg:2011kx, GonzalezSprinberg:2013er}. These are the (a), (d), (e), (f) and (g) diagrams, where we set $\gamma=-\pi/2$ in such a way that the neutral scalar $h$ has the same couplings as the SM Higgs boson. 

\begin{table}[ht]
\centering
\begin{tabular}{|c|c|}
\hline
Type& Particles in \\
&the loop $ABC$\\ \hline
(a)& $\varphi_i\, t\, b$\\ \hline
(b)& $t\, \varphi_i \, H^+$\\ \hline
(c)& $b\, H^+\, \varphi_i$\\ \hline
(d)& $t\, \varphi_i\, G^+$\\ \hline
(e)& $b\, G^+\, \varphi_i$\\ \hline
(f)& $t\, \varphi_i\, W^+$\\ \hline
(g)& $b\, W^+ \, \varphi_i $\\ 
\hline
\end{tabular}
\caption{\label{tabla} Classification of the Feynman diagrams by the type of particles circulating in the loop}
\end{table}

The vertices that contribute to the top anomalous tensor couplings in the A2HDM depend on the scalar mixing angle $\gamma$ and on the alignment parameters $\varsigma_{u}$ and $\varsigma_{d}$. The mass dependence is parameterized by the dimensionless variable $r_{X}=m_{X}/m_{t}$, where $m_X$ is the mass of the particle $X$ circulating in the loop. 
For the neutral scalar masses above the TeV scale the Feynman integrals give negligible values when compared to the EW contributions. However, there is a high sensitivity of the tensor couplings on the masses of the new particles when they take lower values.

\section{Results}\label{Res}

In this section we show the results of our calculation for the top anomalous magnetic moments in the A2HDM. As already stated, the model introduces new physics only in the EW sector. To explicitly show the size of these corrections to the EW sector of the SM, we will compare the new contributions from
the A2HDM  with the values one gets  from the EW contribution of the SM (SM-EW). 
As already stated in the previous paragraph these depend on the masses of the new particles, the alignment parameters $\varsigma_{u,d}$ and the scalar mixing angle $\gamma$. In order to show the new physics effects, we will explicitly present the results as a quotient of the new physics prediction with the SM-EW value for the same anomalous tensor coupling.

We take the current values \cite{Beringer:1900zz} for the standard particles. We chose different sets of values for the masses of the new neutral and charged particles; the different scenarios we consider are shown in Table \ref{escenariosmasas}. The new scalar mass values are taken to be of the order of hundreds of GeV \cite{Aaltonen:2011rj, Abazov:2011ix}. The charged scalar mass, $m_{H^{+}}$, can take values below the top quark mass, so that the decay $t \rightarrow b H^{+}$ is kinematically possible and, therefore, type (c) diagrams may develop an absorptive part. These scenarios are called (\textbf{i}) in our paper and we take for them $m_{H^+}=150$ GeV. For the other cases, where $m_{H^+}>m_t$, we take $m_{H^+}=320$ GeV, as shown in Table \ref{escenariosmasas}. In addition, for a CP conserving scalar potential \cite{Gunion:1989we} we have to impose that $m_{h} \leq m_{H}$. We define six different mass scenarios: two with three light neutral 
scalars (\textbf{I} and \textbf{Ii}), two 
with $h$ as the only light scalar (\textbf{II} and \textbf{IIi}) and two more with the CP-odd scalar $A$ being the lightest one (\textbf{III} and \textbf{IIIi})
\footnote{These scenarios are disfavored by present LHC data
 \cite{Celis:2013rcs}. 
 }. 

\begin{table}[ht]
\begin{center}
\begin{tabular}{|c|c|c|}
\hline 
\multicolumn{3}{|c|}{Scalar mass scenarios (in GeV)}\tabularnewline
\hline 
I & $m_{h}=126$, $m_{H}=173$, $m_{A}=150$, $m_{H^{+}}=320$ & \includegraphics[width=3.3cm,height=0.1cm]{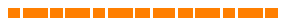}\tabularnewline
\hline 
Ii & $m_{h}=126$, $m_{H}=173$, $m_{A}=150$, $m_{H^{+}}=150$ & \includegraphics[width=3.3cm,height=0.1cm]{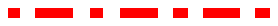}\tabularnewline
\hline 
II & $m_{h}=126$, $m_{H}=865$, $m_{A}=865$, $m_{H^{+}}=320$ & \includegraphics[width=3.3cm,height=0.15cm]{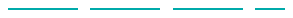}\tabularnewline
\hline 
IIi & $m_{h}=126$, $m_{H}=865$, $m_{A}=865$, $m_{H^{+}}=150$ & \includegraphics[width=3.3cm,height=0.15cm]{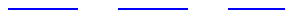}\tabularnewline
\hline 
III & $m_{h}=865$, $m_{H}=865$, $m_{A}=126$, $m_{H^{+}}=320$ & \includegraphics[width=3.3cm,height=0.15cm]{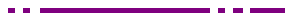}\tabularnewline
\hline
IIIi & $m_{h}=865$, $m_{H}=865$, $m_{A}=126$, $m_{H^{+}}=150$ & \includegraphics[width=3.3cm,height=0.15cm]{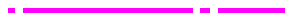}\tabularnewline
\hline 
\end{tabular}
\caption{Different scalar mass scenarios taken for the analysis. Each scenario is identified by a different color and type of line. All masses are in GeV.}
\label{escenariosmasas}
\end{center}
\end{table}
In these last scenarios we take the same mass for both neutral scalar particles ($h$ and $H$). In this case, as expected, the results are independent of the mixing angle $\gamma$ showing that physics cannot separate contributions from degenerate scalar mass-eigenstates. The set of scenarios given in Table \ref{escenariosmasas} allows us to investigate the whole meaningful parameter space and to determine the regions where the tensor couplings take values differing from the SM-EW predictions. In the scenarios (\textbf{II}), (\textbf{IIi}), (\textbf{III}) and (\textbf{IIIi}) the value of the mass of the heaviest scalar or pseudoscalar particle, $865$ GeV, is fixed by setting $r_{heaviest}=(m_{heaviest})/m_t=5$.
We write the alignment parameters as:
\begin{equation}
 \varsigma_{u}=\rho_{u} e^{i\theta_{u}}, \qquad \varsigma_{d}=\rho_{d} e^{i\theta_{d}},
\end{equation}
and we investigate separately the effects of modulus and phases on the anomalous tensor couplings $ g_{R,L}$. 
Besides the masses of the new particles we have five free parameters: $\rho_{u}$, $\rho_{d}$, $\theta_{u}$, $\theta_{d}$ and the mixing angle $\gamma$.

The deviations from the predictions of the EW sector of the SM are shown using the ratios:

\begin{equation}
Q^{Re}_{R}\equiv \frac{Re\left(g^{A2HDM}_{R}\right)}{Re\left(g^{EW}_{R}\right)},\quad Q^{Re}_{L}\equiv \frac{Re\left(g^{A2HDM}_{L}\right)}{Re\left(g^{EW}_{L}\right)}, 
\end{equation}

\begin{equation}
Q^{Im}_{R}\equiv \frac{Im\left(g^{A2HDM}_{R}\right)}{Im\left(g^{EW}_{R}\right)},\quad Q^{Im}_{L}\equiv \frac{Im(g^{A2HDM}_{L})}{Im(g^{EW}_{L})}. 
\end{equation} 
These are the quotients of the real and imaginary parts of the tensor couplings calculated in the A2HDM ($g_{R,L}^{A2HDM}$) and the EW contributions of the SM to them ($g_{R,L}^{EW}$), given in Eq. (\ref{ew-contr}).

For the six different mass scenarios defined in Table \ref{escenariosmasas}, we will show how these quotients depend on the four alignment parameters $\rho_{d, u}$, $\theta_{u,d}$, and on the mixing angle $\gamma$. 

In general, we will show the results for conservative values of the modulus, {\it i.e.} for $\rho_{u,d} \sim 1$. For greater values of the modulus will certainly produce larger deviations from the SM-EW predictions.

\subsection{Contributions to \texorpdfstring{$g_{R}$}{gR}}\label{cgr}

As a general feature we found that even for values of the modules $\rho_{u, d} \simeq 1$, the values of the anomalous tensor couplings in the A2HDM can be significantly different from the SM-EW values. We also found that the right coupling $ g_{R}$ is independent of the down-type alignment parameter $\varsigma_{d}$ for the same range of values as we take for $\varsigma_{u}$; this is due to the fact that diagrams with top quarks in the loop, specially types (b) and (d), give a big numerical contribution because the top Yukawa vertex is proportional to $\varsigma_{u}$ and to $m_t$. In order to fix values, we take $\rho_d = 1$ and $\theta_d = \pi / 4 $ to analyze the dependence of $Q^{Re}_{R}$ and $Q^{Im}_{R}$ on the other parameters.

\begin{figure}[h]
 \centering
 \subfloat[]{\label{QRegRg1}\includegraphics[width=0.49\textwidth]{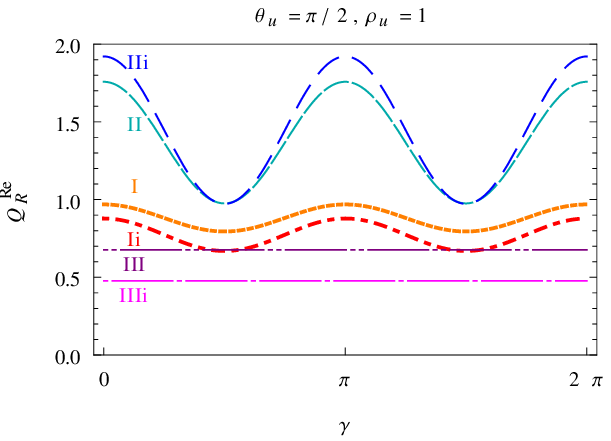}}
 ~
 \subfloat[]{\label{QRegRg2}\includegraphics[width=0.49\textwidth]{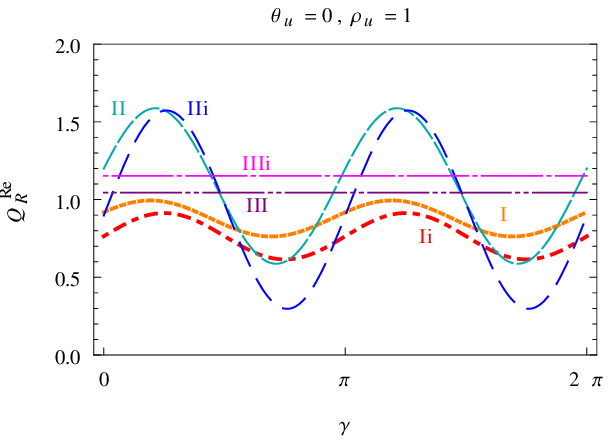}}
 \caption{Plot of the quotient $Q^{Re}_{R}$ as a function of $\gamma$ for the different mass scenarios.}
 \label{RegRgamma}
 \end{figure}

In Fig.\ \ref{RegRgamma} we plot the dependence of the quotient $Q^{Re}_{R}$ on the scalar mixing angle $\gamma$ for the selected values of $\theta_u$ and $\rho_u$. It shows a smooth variation for scenarios (\textbf{I}) and (\textbf{Ii}), and a stronger dependence for scenarios (\textbf{II}) and (\textbf{IIi}). As already mentioned, we found that in scenarios (\textbf{III}) and (\textbf{IIIi}) there is no dependence on $\gamma$. This is due to the fact that the two neutral scalars are degenerate in these scenarios and, then, the mixing angle has no physical content. 

\begin{figure}[h]
 \centering
 \subfloat[]{\label{QImgRg1}\includegraphics[width=0.49\textwidth]{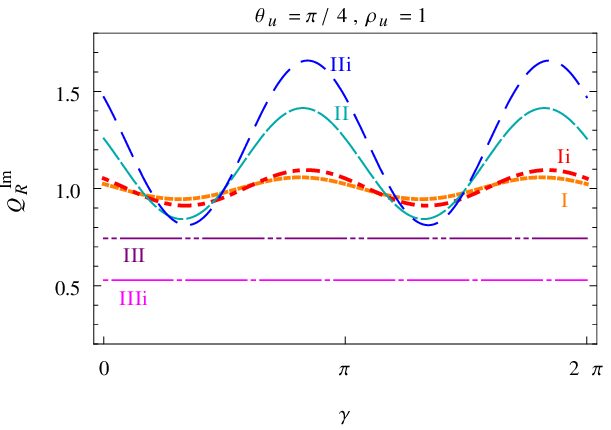}}
 ~
 \subfloat[]{\label{QImgRg2}\includegraphics[width=0.49\textwidth]{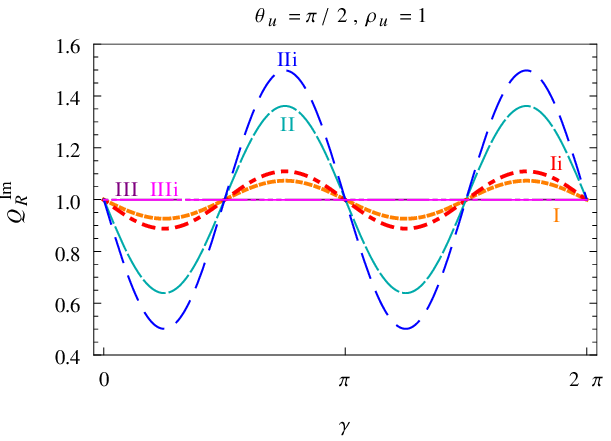}}
 \caption{Plot of the quotient $Q^{Im}_{R}$ as a function of $\gamma$ for the different mass scenarios. }
 \label{QImgRg}
 \end{figure}

The real and imaginary parts of $g_R$ have, in general, deviations from the SM-EW values of the same order of magnitude. The dependence of $Q^{Im}_{R}$ on the mixing angle $\gamma$ is shown in Fig.\ \ref{QImgRg}. We show the plots for $\theta_{u}= \pi/4$ and $\theta_{u}= \pi/2$ where the mass scenarios (\textbf{II}) and (\textbf{IIi}) have great sensitivity to the angle $\gamma$. 

\begin{figure}[h]
 \centering
 \subfloat[]{\label{QRegRtu}\includegraphics[width=0.49\textwidth]{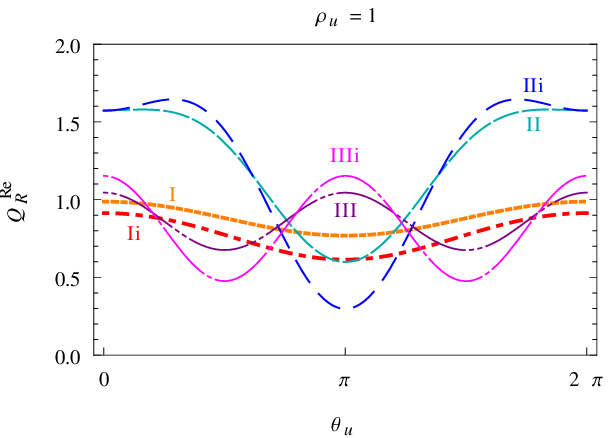}}
 ~
 \subfloat[]{\label{QImgRtu}\includegraphics[width=0.49\textwidth]{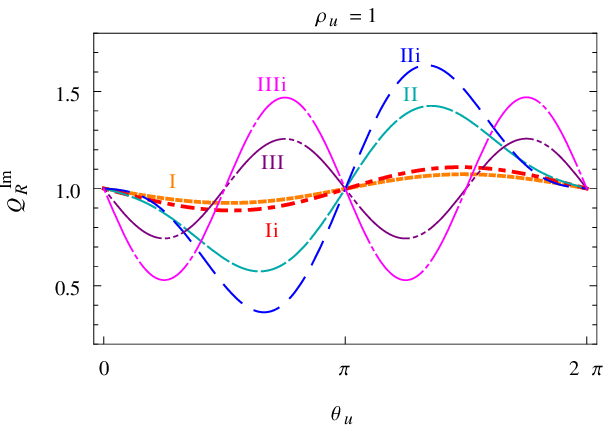}}
 \caption{Plot of the quotients $Q^{Re}_{R}$ and $Q^{Im}_{R}$ as a function of $\theta_{u}$ for the different mass scenarios. We take $\gamma = \pi/4$.}
 \label{thu}
 \end{figure}

In Fig.\ \ref{QRegRtu} we plot $Q^{Re}_{R}$ as a function of the phase $\theta_{u}$. This plot, and the ones that follows, do not strongly depend on the value of $\gamma$, and we choose $\gamma=\pi/4$ for all of them.
We found a strong dependence on the phase for scenarios with big scalars mass differences, being the most sensitive the ones labeled (\textbf{II}) and (\textbf{IIi}). In the last one we found that $Q^{Re}_{R}$ can even take negative values for $\theta_{u}\approx \pi$ when increasing the value of $\rho_{u}$ up to $2$. This means that $Re(g_{R})$ can take positive values, contrary to the negative sign one gets in the SM-EW prediction. The dependence of $Q^{Im}_{R}$ with $\theta_{u}$ is shown in Fig.\ \ref{QImgRtu}, where again mass scenarios (\textbf{II}), (\textbf{IIi}), and (\textbf{III}), (\textbf{IIIi}) are the most sensitive ones. In the same figure, it can be seen that if we consider the alignment parameter $\varsigma_{u}$ to be real and $\rho_{u}=1$, the value of $Im(g_{R})$ is the same as the one given by the EW sector of the SM, for all mass scenarios. Finally, we also obtained that when the alignment parameter $\varsigma_u$ is real, the 
dependence of $Q^{Im}_{R}$ with the modulus $\rho_u$ is almost 
negligible.

\begin{figure}[h]
 \centering
 \subfloat[]{\label{QRegRrou1}\includegraphics[width=0.49\textwidth]{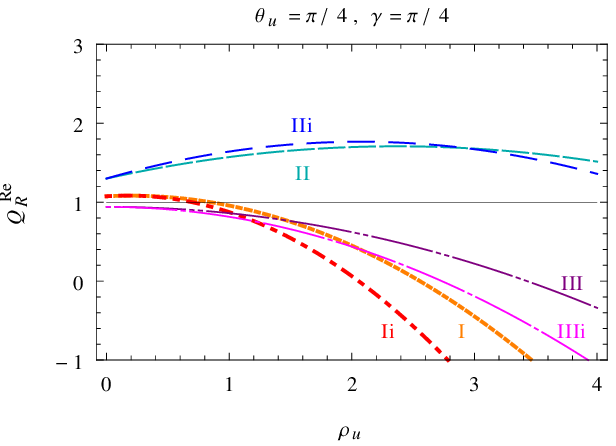}}
 ~ 
 \subfloat[]{\label{QRegRrou3}\includegraphics[width=0.49\textwidth]{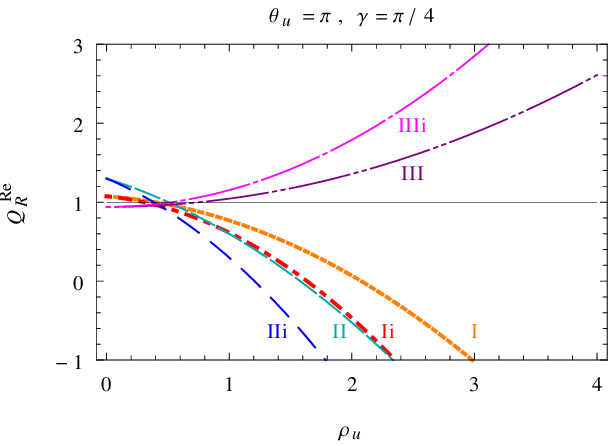}}
 \caption{Plot of the quotient $Q^{Re}_{R}$ as a function of $\rho_{u}$, for different mass scenarios and phases of the alignment parameters.}
 \label{QRegRrou}
 \end{figure}

Figure \ref{QRegRrou} shows the dependence of $Q^{Re}_{R}$ with the modulus $\rho_{u}$ for all the mass scenarios we considered and for different values of $\theta_{u}$. Keeping $\rho_{u}\lesssim 1.5$ it can be checked that, for all phases and mass scenarios, $Re(g_{R})$ has the same sign as in the SM. 
For bigger values of $\rho_u$, deviations from the SM-EW value, in general, grow as $\rho_u$ increases, but the particular behavior depends on the set of masses chosen. By the two examples shown in Fig. \ref{QRegRrou}, it can be seen that large deviations from the SM-EW prediction (for example, bigger than 50\%) can be found in almost all scenarios for $1<\rho_u <2$ and for any choice of the phase $\theta_u$.

\begin{figure}[h]
 \centering
 \subfloat[]{\label{QImgRrou1}\includegraphics[width=0.49\textwidth]{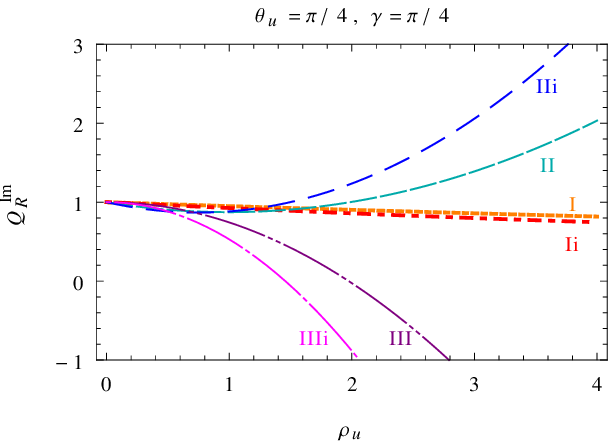}}
 ~
 \subfloat[]{\label{QImgRrou2}\includegraphics[width=0.49\textwidth]{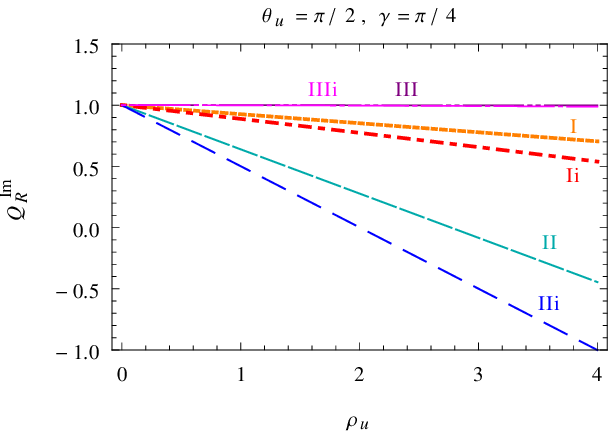}}
 \caption{Plot of the quotient $Q^{Im}_{R}$ as a function of $\rho_{u}$, for different mass scenarios and phases of the alignment parameters.}
 \label{QImgRrou}
 \end{figure}

Figure \ref{QImgRrou} shows the dependence of $Q^{Im}_{R}$ on $\rho_{u}$. We found that the $Q^{Im}_{R}$ plots, as a function of $\rho_u$, are symmetric with respect to the $Q^{Im}_{R}=1 $ line when changing the phase from $\theta_u$ to $(2\pi-\theta_u)$. 
In Fig. \ref{QImgRrou1} ($\theta_u=\pi/4$) and in Fig. \ref{QImgRrou2} ($\theta_u=\pi/2$) we show that an important deviation from the EW value is obtained. However, for $\theta_u=0$ 
we found a small deviation of less than five per mil from the values of the SM-EW prediction for $Im(g_{R})$. This originates in the smallness of the new absorptive
parts coming from the non standard contributions that are still present for $\theta_u = 0$. We also found that the deviation from the SM-EW calculation grows with $\rho_{u}$ for almost all mass scenarios and for $\theta_u \neq 0$. 
For $\theta_u=\pi/2$ and scenarios (\textbf{II}) and (\textbf{IIi}) this deviation can be very 
strong while it is negligible for scenarios (\textbf{III}) and (\textbf{IIIi}). Even if $1 <\rho_u < 2$ we still found that some scenarios 
(types (\textbf{II}) and (\textbf{IIi}), for $\theta=\pi/2$, and types (\textbf{III}) and (\textbf{IIIi}), for $\theta=\pi/4$, for example) show a strong departure from the SM-EW value.
We can see in Fig. \ref{QImgRtu} and Fig. \ref{QImgRrou} that, taking into account the current bounds on $\rho_{u}$, provided by flavor physics ($\rho_{u}<2$), we still find a sizable deviation in the predicted value for $Im(g_R)$ and, therefore, for the normal asymmetry, $A^{N}_{FB}$.
This result would point to a non zero complex phase $\theta_{u}$ and would exclude some of the mass scenarios selected.

In general, one always can find specific mass scenarios and phases that produce sizable deviations of $Im(g_R)$ from the SM-EW value (for example, bigger than 50\%). This is not the case only for $\theta_u=0$ where there is almost no deviation from the SM-EW prediction.

\subsection{Contributions to \texorpdfstring{$g_L$}{gL}}\label{cgl}

\begin{figure}[h]
 \centering
 \subfloat[]{\label{QImgLg1}\includegraphics[width=0.5\textwidth]{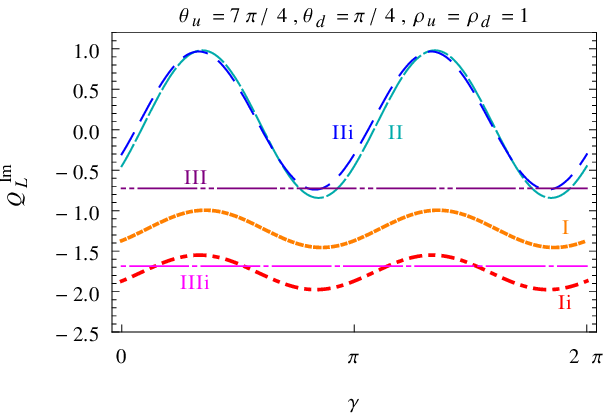}} ~
 \subfloat[]{\label{QImgLg2}\includegraphics[width=0.486\textwidth]{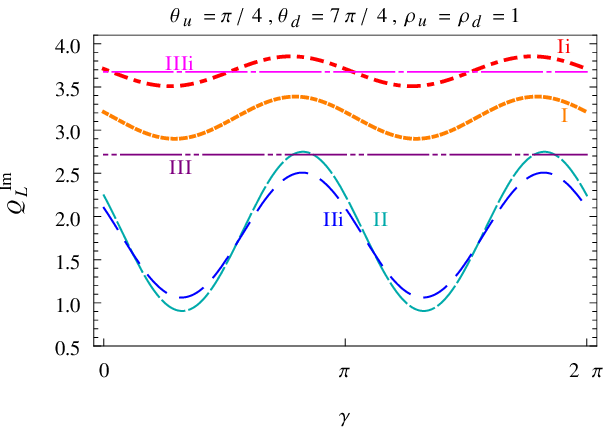}}
 \caption{Plot of the quotient $Q^{Im}_{L}$ as a function of $\gamma$ for the different mass scenarios and phases 
 $\theta_{u,d}$.}
 \label{RegLgamma}
 \end{figure}

The left tensor coupling $ g_{L}$ depends on both alignment parameters $\varsigma_{u}$ and $\varsigma_d$. The quotient $Q^{Re}_{L}$ shows a soft dependence on the scalar mixing angle $\gamma$ for all the mass scenarios considered, and for every $\theta_{u,d}$ combination explored. In the scenarios (\textbf{III}) and (\textbf{IIIi}) we found that, as it was the case for $g_R$, there is no $\gamma$-dependence at all. The oscillation of $Im(g_{L})$ with the mixing angle $\gamma$ is maximum when we take the phases $\theta_u + \theta_d = 2 \pi$. As an example, the cases ($\theta_{u}=7\pi/4 , \theta_{d}=\pi/4$) and ($\theta_{u}=\pi/4, \theta_{d}=7\pi/4$) are shown in Fig.\ \ref{RegLgamma}.

\begin{figure}[h]
 \centering
 \subfloat[]{\label{QRegLrou1}\includegraphics[width=0.49\textwidth]{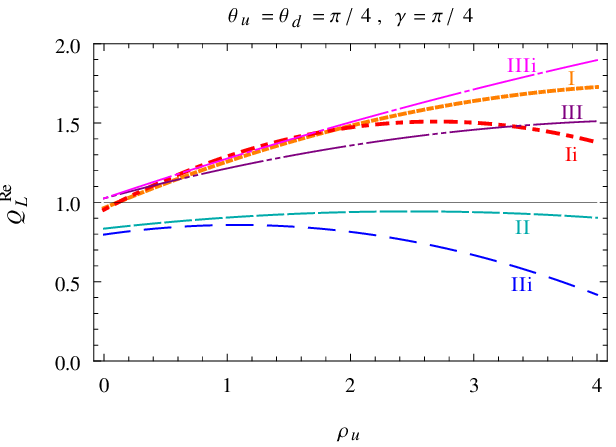}}
 ~
 \subfloat[]{\label{QRegLrou2}\includegraphics[width=0.49\textwidth]{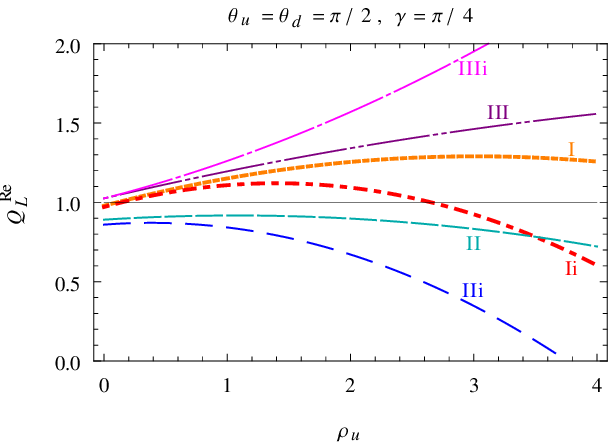}}
 
 \subfloat[]{\label{QRegLrou3}\includegraphics[width=0.49\textwidth]{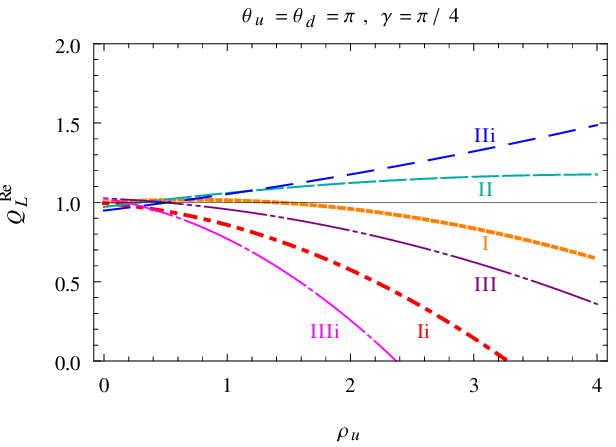}}
 ~
 \subfloat[]{\label{QRegLrou4}\includegraphics[width=0.49\textwidth]{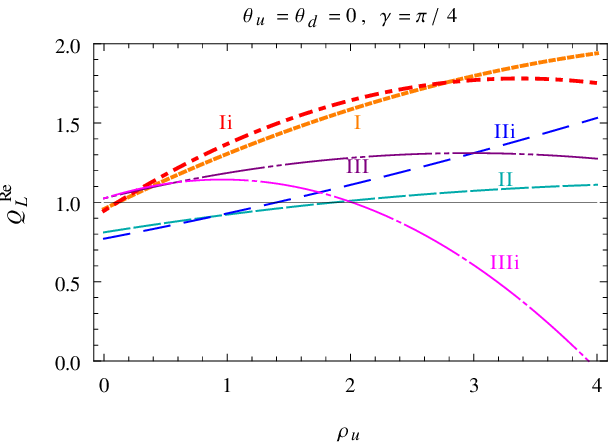}}
 \caption{Plot of the quotient $Q^{Re}_{L}$ as a function of $\rho_{u}$, for different mass scenarios and phases of the alignment parameters.}
 \label{QRegLrou}
 \end{figure}

 The real part of the $g_L$ coupling shows a strong dependence on the $\rho_{u}$ parameter, while the dependence on the phases $\theta_{u}$ and $\theta_d$ is softer. This is shown in Fig. \ref{QRegLrou} for different values of $\theta_{u,d}$.

As can be seen from Fig.\ref{QRegLrou} an appreciable deviation of $Re( g_{L})$ from the SM-EW value needs, in general, large values of the $\rho_u$ parameter, such as $\rho_{u}>2$. The quotient $Q^{Re}_{L}$ is always positive if $\rho_{u}<2$. The current SM prediction for $Re( g_{L})$ \cite{GonzalezSprinberg:2011kx, GonzalezSprinberg:2013er} is slightly below the lower bound suggested in Ref. \cite{Drobnak:2011aa}. This means that a positive contribution is needed in order to fit this limit. We found that the A2HDM can only accommodate that situation by taking values of $\rho_{u}>2$, in tension with the current bounds given by flavor physics \cite{Jung:2010ik}.
Only for selected phases and mass scenarios ($\theta_{u,d}=\pi$ for types (\textbf{Ii}) or (\textbf{IIIi}), and $\theta_{u,d}=0$ for types (\textbf{I}) or (\textbf{Ii}), for example) one can have deviations of the order of 50\%, for $\rho_u \sim 2$.
 
We have also considered the dependence of $Q^{Re}_{L}$ with respect to $\rho_d$, but no important changes result from this dependence. The quotient $Q^{Re}_{L}$ has a linear dependence with $\rho_d$. For example, keeping
$\theta_u = \theta_d$ as $\rho_d$ grows, $Q^{Re}_{L}$ exhibits a linear growing with a positive slope lower than one for every mass scenario. No important consequences are found for $Re(g_L)$ in this case and, for example, no change in the sign of $Re(g_L)$ is found with respect to the SM-EW
prediction. If the phases $\theta_u$ and $\theta_d$ are taken to be in the intervals $(0,\pi)$
and $(\pi, 2 \pi)$, respectively, the quotient $Q^{Re}_{L}$ decreases but is always positive for
every mass scenario for $\rho_d < 4$.

\begin{figure}[h]
 \centering
 \subfloat[]{\label{QImgLtu}\includegraphics[width=0.5\textwidth]{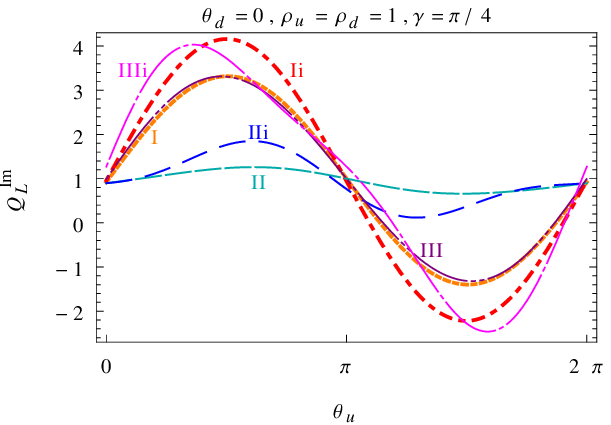}}
 ~
 \subfloat[]{\label{QImgLtd}\includegraphics[width=0.486\textwidth]{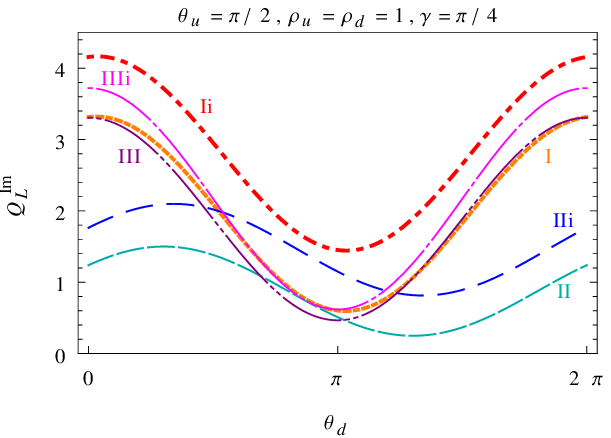}}
 \caption{Plot of the quotient $Q^{Im}_{L}$ as a function of $\theta_{u}$ and $\theta_{d}$ for the different mass 
 scenarios.}
 \label{QImgLt}
 \end{figure}

Figure \ref{QImgLt} shows the strong dependence of $Q^{Im}_{L}$ on the phases $\theta_{u,d}$. There it can also be seen that, due to the fact that the value of $Im( g_{L})$ is very small in the SM, the quotient $Q^{Im}_{L}$ is very sensitive to the $\theta_{u,d}$ phases and to the mixing angle $\gamma$. This magnitude, if measured, may allow a clear distinction between scenarios (\textbf{II, IIi}) and the rest of them. 

\begin{figure}[h]
 \centering
 \subfloat[]{\label{QImgLrou1}\includegraphics[width=0.49\textwidth]{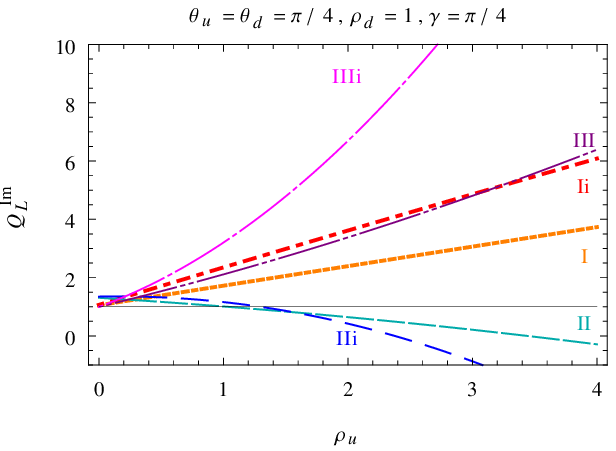}}
 ~
 \subfloat[]{\label{QImgLrou2}\includegraphics[width=0.49\textwidth]{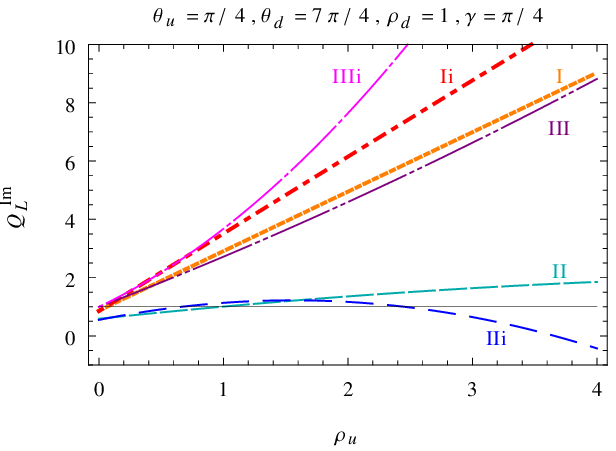}}
 \caption{Plot of the quotient $Q^{Im}_{L}$ as a function of $\rho_{u}$, for different mass scenarios and phases of the alignment parameters.}\label{QImgLrou}
 \end{figure}

The $Q^{Im}_{L}$ dependence on $\rho_{u}$ is shown in Fig.\ \ref{QImgLrou}, for different $\theta_{u,d}$ phases and taking $\rho_d=1$. While for mass scenarios (\textbf{I}) and(\textbf{Ii}) it exhibits a linear dependence for the whole scanned $\rho_{u}$ range, its dependence for scenarios (\textbf{II}), (\textbf{IIi}) and (\textbf{III}), (\textbf{IIIi}) is quadratic.

For $\theta_{u}= 7\pi/4$ the plots are symmetric, with respect to the $Q^{Im}_{L}=1 $ line, to the ones shown in Fig.\ \ref{QImgLrou}. In addition, as can be seen there, deviations from the SM-EW values bigger than 100\% are found for low values of the $\rho_u$ parameter ({\it i.e.}, $\rho_u<2$).
On the other hand, we have checked that $Im( g_{L})$ is almost independent on $\rho_{u}$ when we choose the alignment parameters to be real. In that case, the only scenarios where $Im( g_{L})$ shows deviations from the SM value greater than 100\%
for $\rho_{u}<3$ are types (\textbf{IIi}) and (\textbf{IIIi}). This is again due to new absorptive parts that are not present in the SM.

In all the mass scenarios, for $\rho_u=1$ and for any value of the phases, we found that $Q^{Im}_{L}$ grows linearly with $\rho_{d}$ with positive slopes up to $1$. Similarly as stated for $Im(g_R)$, a sizable deviation in the predicted value of the $W$ helicity fraction $\rho_+\simeq F_+/F_0$ \cite{AguilarSaavedra:2007rs, AguilarSaavedra:2010nx} would point to non zero complex phases $\theta_{u,d}$.

\subsection{Contributions from Type I and II 2HDM}\label{IyII}

\begin{figure}[h]
 \centering
 \subfloat[]{\label{ReGLtI}\includegraphics[width=0.49\textwidth]{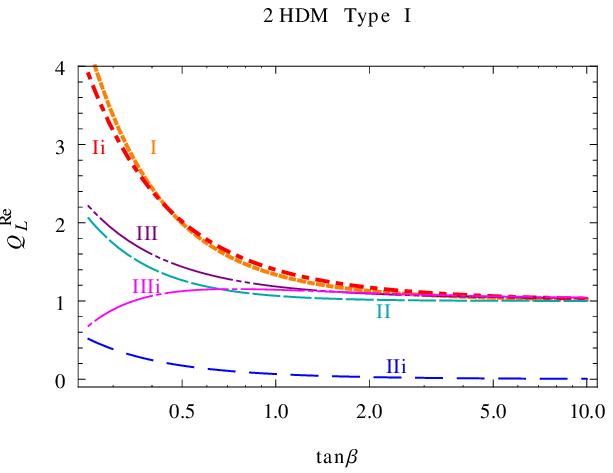}}
 ~
 \subfloat[]{\label{ImGLtI}\includegraphics[width=0.49\textwidth]{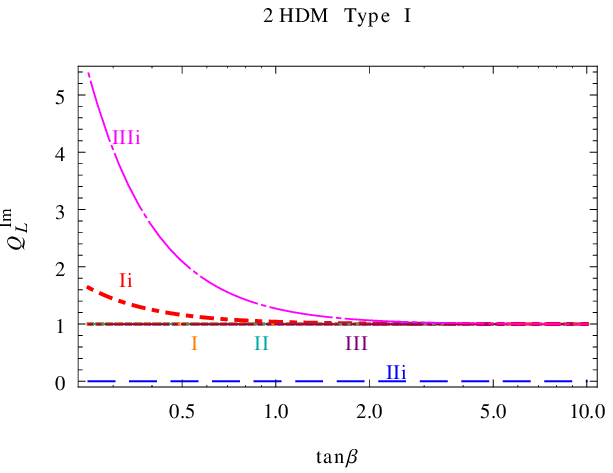}}
 
 \subfloat[]{\label{ReGRtI}\includegraphics[width=0.49\textwidth]{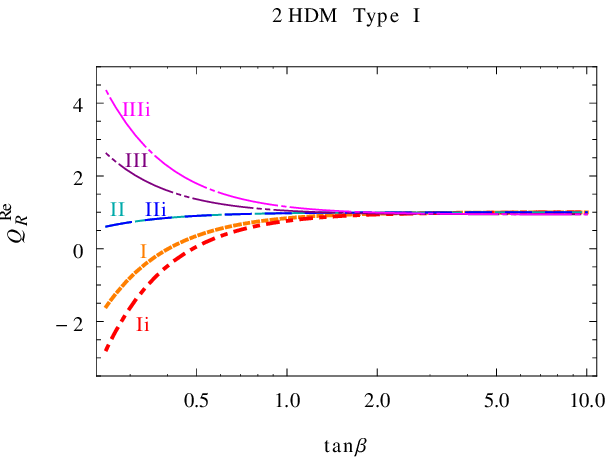}}
 ~
 \subfloat[]{\label{ImGRtI}\includegraphics[width=0.49\textwidth]{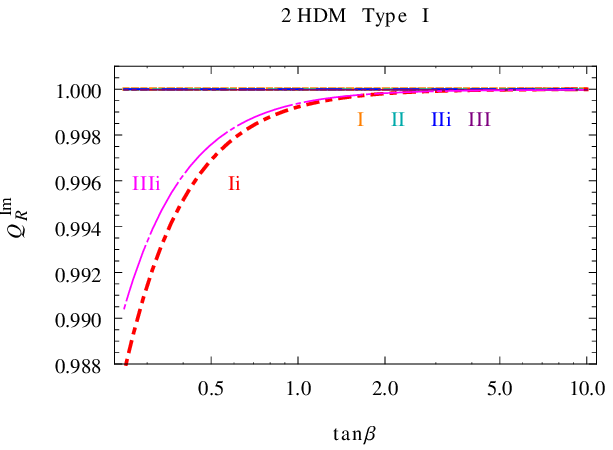}}
 \caption{Plot of the quotients $Q^{Re}_{L}$, $Q^{Im}_{L}$, $Q^{Re}_{R}$ and $Q^{Im}_{R}$ as a function of $\tan\beta$ for the different mass scenarios in Type I 2HDM and $\beta = \alpha + \pi/2$.}
 \label{gLRTI}
 \end{figure}

\begin{figure}[h]
 \centering
 \subfloat[]{\label{ReGLtII}\includegraphics[width=0.49\textwidth]{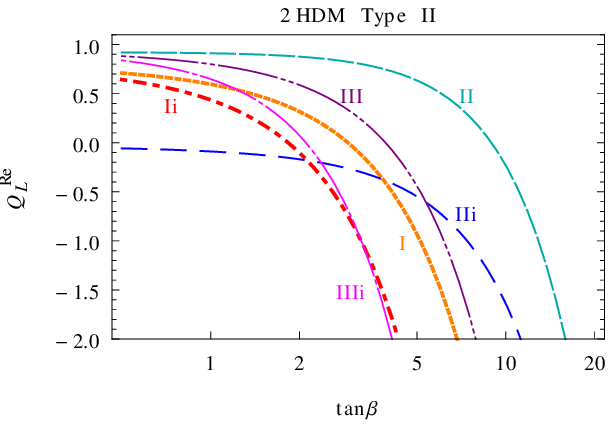}}
 ~
 \subfloat[]{\label{ImGLtII}\includegraphics[width=0.49\textwidth]{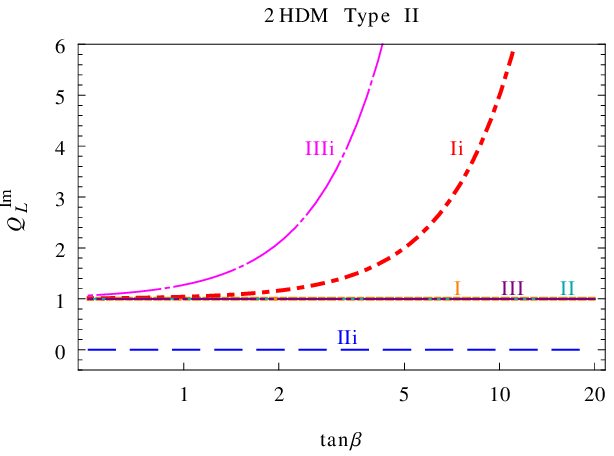}}
 
 \subfloat[]{\label{ReGRtII}\includegraphics[width=0.49\textwidth]{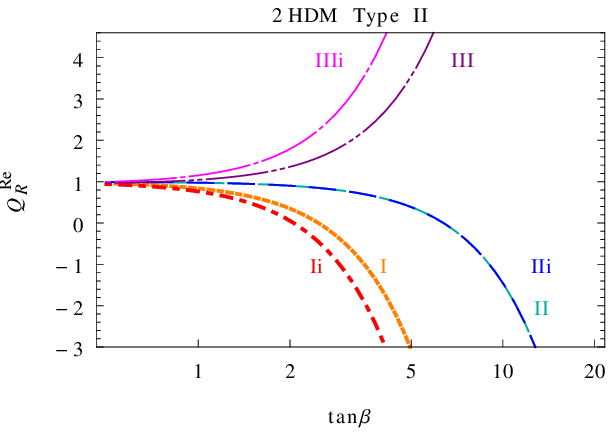}}
 ~
 \subfloat[]{\label{ImGRtII}\includegraphics[width=0.49\textwidth]{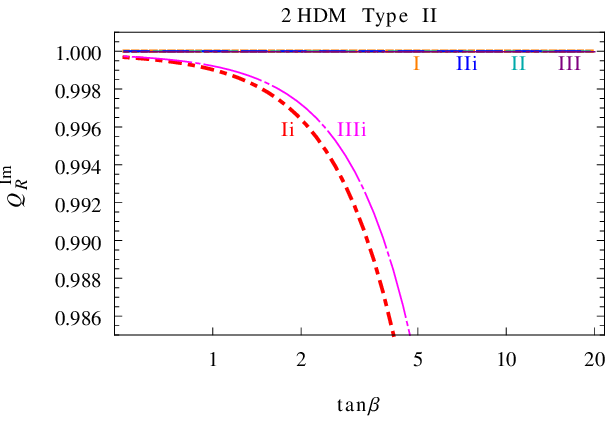}}
 \caption{Plot of the quotients $Q^{Re}_{L}$, $Q^{Im}_{L}$, $Q^{Re}_{R}$ and $Q^{Im}_{R}$ as a function of $\tan\beta$ for the different mass scenarios in Type II 2HDM and $\beta=\alpha + \pi /2$.}
 \label{gLRTII}
 \end{figure}

For the particular case of the Type I and II 2HDM we found, taking the appropriate limit in the A2HDM parameters, the results that are shown in Fig. \ref{gLRTI} and \ref{gLRTII}. As previously mentioned, Types I and II 2HDM can be recovered by choosing real alignment parameters $\varsigma_{u,d}$ \cite{Pich:2009sp}. The couplings in these models are usually written in a generic basis where both Higgs doublets acquire a vacuum expectation value. Then, the key parameters are the ratio of these values, parameterized as $\tan\beta=v_{2}/v_{1}$ and the mixing angle, $\alpha$, between the two CP-even neutral scalars in this basis. For a CP-conserving potential we choose $\tan\beta$ and the mixing angle of the two neutral scalars $h$ and $H$ in such a way that $\beta = \alpha + \pi/2$. Therefore, the neutral scalar $h$ has SM-like coupling to the photon and to the weak bosons.

Taking the alignment parameters to be $\varsigma_{u}=\cot\beta$ and $\varsigma_{d}=\cot\beta$, one recovers the Type I 2HDM, shown in Fig. \ref{gLRTI}. As expected, for low values of $\tan\beta$ the real parts show important deviations from the SM-EW values, and a similar behavior is found for $Q^{Im}_{L}$. The variation for $Q^{Im}_{R}$ is negligible even for the type (\textbf{i}) scenarios.

Setting $\varsigma_{u} = \cot\beta$ and $\varsigma_{d}= - \tan\beta$ one recovers the Type II 2HDM. In Fig. \ref{gLRTII} we show our results for this model. As expected, the limit $\tan\beta \gg 1$ gives singular results for the real parts in all the mass scenarios, while the imaginary part is singular in this limit only in the (\textbf{Ii}) and (\textbf{IIIi}) scenarios. The most important deviation from the SM predictions is found for $Q^{Im}_{L}$, in scenario (\textbf{IIIi}), dominated by the contribution of the pseudo-scalar $A$.

\section{Conclusions}\label{Conc}

We have calculated the contributions to the  top anomalous tensor couplings $ g_{R,L}$ in the A2HDM with a CP-conserving potential. We have compared the numerical predictions of the model with the electroweak SM values for different scalars mass scenarios. The complete values of the couplings can be obtained by adding the QCD contribution given in Eq. (\ref{qcd-contr}), to the calculated in this paper
in the A2HDM.

The parameter space of the model has been extensively explored. There are large regions of this space where important deviations of the top tensor couplings from the predictions of the EW sector of the SM can be found.
 The four couplings $Re(g_R)$, $Im(g_R)$, $Re(g_L)$ and $Im(g_L)$ show a remarkable sensitivity to the new physics parameters and in extended regions of this parameter space they have large deviations from the EW predictions.

The study of the A2HDM in this unexplored context of top quark physics shows that the precise measurement of these magnitudes may allow for a discrimination among the different scalar mass scenarios and the value of the mixing angle $\gamma$. The measurement of the anomalous tensor couplings (real and imaginary parts) for the top quark can also reveal new CP-violation mechanisms that can be accounted for in the A2HDM by the the complex alignment parameters $\varsigma_{u,d}$. The observables considered in the literature, taken together with the results presented in this paper can help in finding new physics and also in restricting the range of the allowed regions of the parameter space in the A2HDM.

As expected, the right coupling $g_{R}$ is not sensitive to the down alignment parameter $\varsigma_{d}$. The dependence of the real part $Re( g_{R})$ with $\rho_{u}$ allows to discri\-mi\-na\-te between different scalar mass scenarios. Taking into account the constraint $\rho_{u} < 2$, coming from B physics, one can also have deviations from the SM-EW prediction up to 100\%.  This means that the contribution of the A2HDM to $Re( g_{R})$ can, by itself, make this number a  15\% higher than the SM prediction.
These large effects can even change the sign of the A2HDM contribution to the  top couplings with respect to the electroweak SM prediction for different values of $\theta_{u}$. Despite not being the most sensitive quantity to the value of the new physics parameters, the quotient $Q^{Im}_{R}$ can take values from 0.5 up to 1.5, for $\rho_u=1$ and for several values of $\theta_{u}$. We also have found that the absorptive parts are less than 5\% of the electroweak SM value even for high values of $\rho_{u}$ ({\it i.e.} $\rho_u\simeq 4$), for a pure real alignment parameter $\varsigma_{u}$. $Im(g_R)$ has already been 
measured at the LHC \cite{ATLAS-CONF-2013-032}, taking advantage of the recently investigated asymmetries in the normal direction \cite{AguilarSaavedra:2010nx}, and future measurements may show sensitivity to new physics. A significant deviation of this measurement from the electroweak SM value would point to new CP-violation mechanisms such as the non-zero phases $\theta_{u,d}$. 

The left anomalous tensor coupling $ g_{L}$ shows sizable dependencies on both alignment parameters, and both the real and imaginary parts are very sensitive to these parameters. For $Re(g_L)$ there are some values of the $\rho_u$ parameter where this magnitude can change sign with respect to the electroweak SM prediction. Then, the total one loop QCD plus A2HDM prediction for this coupling  is 18\% lower than the SM one given in Eq.(\ref{sm-contr}).  Besides, this  fact could produce contributions that may elucidate the tension between the indirect bounds put on $Re( g_{L})$ by $b\rightarrow s \gamma$ decays and the SM prediction. The imaginary part of $ g_{L}$ is extremely sensitive to $\rho_{u}$ and to both complex phases $\theta_{u,d}$. We have found that it can deviate from the SM prediction up to 400\%, even for low values of the $\rho_{u}$ parameter ($\simeq 1$).

In both Type I and II models we found that the real parts have important deviations from the EW values. The imaginary parts of the couplings also have sizable deviations from the electroweak SM predictions in some of the studied mass scenarios.
 For the Type I models and for low values of $\tan\beta$ we found that the real parts deviate strongly from the EW prediction, while this only occurs in some of the scenarios for the imaginary part. The limit $\tan\beta \gg 1$ gives singular results for the real parts in Type II models in all the scenarios, while the imaginary part is singular in this limit in scenarios (\textbf{Ii}) and (\textbf{IIIi}).

High precision measurements of the top quark anomalous tensor couplings are expected in the next high energy runs at the LHC and in the next generation of colliders. These measurements, the flavor constraints and the collider searches for new scalar resonances are complementary insights and will illuminate this up to now almost unexplored physics. 

\section{Acknowledgments}

This work has been supported, in part, by the Ministerio de Ciencia e Innovaci\'on, Spain, under grants FPA2011-23897 and FPA2011-23596; by Generalitat Valenciana, Spain, under grant PROMETEO/2009/128; and by Pedeciba, CSIC, and ANII under grant BE-POS-2010-2260 and PR-FCE-2009-1-2986, Uruguay.

\bibliographystyle{biblio-estilo}
\bibliography{BibLu_mod}

\begin{thebibliography}{10}%
\makeatletter
\providecommand \@ifxundefined [1]{%
 \ifx #1\undefined \expandafter \@firstoftwo
 \else \expandafter \@secondoftwo
\fi
}%
\providecommand \@ifnum [1]{%
 \ifnum #1\expandafter \@firstoftwo
 \else \expandafter \@secondoftwo
\fi
}%
\providecommand \enquote [1]{``#1''}%
\providecommand \bibnamefont  [1]{#1}%
\providecommand \bibfnamefont [1]{#1}%
\providecommand \citenamefont [1]{#1}%
\providecommand\href[0]{\@sanitize\@href}%
\providecommand\@href[1]{\endgroup\@@startlink{#1}\endgroup\@@href}%
\providecommand\@@href[1]{#1\@@endlink}%
\providecommand \@sanitize [0]{\begingroup\catcode`\&12\catcode`\#12\relax}%
\@ifxundefined \pdfoutput {\@firstoftwo}{%
 \@ifnum{\z@=\pdfoutput}{\@firstoftwo}{\@secondoftwo}%
}{%
 \providecommand\@@startlink[1]{\leavevmode\special{html:<a href="#1">}}%
 \providecommand\@@endlink[0]{\special{html:</a>}}%
}{%
 \providecommand\@@startlink[1]{%
  \leavevmode
  \pdfstartlink
   attr{/Border[0 0 1 ]/H/I/C[0 1 1]}%
   user{/Subtype/Link/A<</Type/Action/S/URI/URI(#1)>>}%
  \relax
 }%
 \providecommand\@@endlink[0]{\pdfendlink}%
}%
\providecommand \url  [0]{\begingroup\@sanitize \@url }%
\providecommand \@url [1]{\endgroup\@href {#1}{\urlprefix}}%
\providecommand \urlprefix [0]{URL }%
\providecommand \Eprint[0]{\href }%
\@ifxundefined \urlstyle {%
  \providecommand \doi [1]{doi:\discretionary{}{}{}#1}%
}{%
  \providecommand \doi [0]{doi:\discretionary{}{}{}\begingroup
  \urlstyle{rm}\Url }%
}%
\providecommand \doibase [0]{http://dx.doi.org/}%
\providecommand \Doi[1]{\href{\doibase#1}}%
\providecommand \bibAnnote [3]{%
  \BibitemShut{#1}%
  \begin{quotation}\noindent
    \textsc{Key:}\ #2\\\textsc{Annotation:}\ #3%
  \end{quotation}%
}%
\providecommand \bibAnnoteFile [2]{%
  \IfFileExists{#2}{\bibAnnote {#1} {#2} {\input{#2}}}{}%
}%
\providecommand \typeout [0]{\immediate \write \m@ne }%
\providecommand \selectlanguage [0]{\@gobble}%
\providecommand \bibinfo [0]{\@secondoftwo}%
\providecommand \bibfield [0]{\@secondoftwo}%
\providecommand \translation [1]{[#1]}%
\providecommand \BibitemOpen[0]{}%
\providecommand \bibitemStop [0]{}%
\providecommand \bibitemNoStop [0]{.\EOS\space}%
\providecommand \EOS [0]{\spacefactor3000\relax}%
\providecommand \BibitemShut [1]{\csname bibitem#1\endcsname}%
\bibitem{:2012gk}%
  \BibitemOpen
  \bibfield{author}{%
  \bibinfo {author} {\bibfnamefont{G.}~\bibnamefont{Aad}} \emph{et~al.}
  (\bibinfo {collaboration} {ATLAS Collaboration}),\ }%
  \emph{\bibinfo {title} {{Observation of a new particle in the search for the
  Standard Model Higgs boson with the ATLAS detector at the LHC}}},\
  \bibfield{journal}{%
  \bibinfo {journal} {Phys. Lett. B}}%
   (\bibinfo {year} {2012}),\
  \Eprint{http://arxiv.org/abs/1207.7214}{arXiv:1207.7214 [hep-ex]}.%
  \bibAnnoteFile{Stop}{:2012gk}%
\bibitem{:2012gu}%
  \BibitemOpen
  \bibfield{author}{%
  \bibinfo {author} {\bibfnamefont{S.}~\bibnamefont{Chatrchyan}} \emph{et~al.}
  (\bibinfo {collaboration} {CMS Collaboration}),\ }%
  \emph{\bibinfo {title} {{Observation of a new boson at a mass of 125 GeV with
  the CMS experiment at the LHC}}},\ \bibfield{journal}{%
  \bibinfo {journal} {Phys. Lett. B}}%
   (\bibinfo {year} {2012}),\
  \Eprint{http://arxiv.org/abs/1207.7235}{arXiv:1207.7235 [hep-ex]}.%
  \bibAnnoteFile{Stop}{:2012gu}%
\bibitem{Abazov:2010jn}%
  \BibitemOpen
  \bibfield{author}{%
  \bibinfo {author} {\bibfnamefont{V.~M.}\ \bibnamefont{Abazov}} \emph{et~al.}
  (\bibinfo {collaboration} {D0 Collaboration}),\ }%
  \emph{\bibinfo {title} {{Measurement of the W boson helicity in top quark
  decays using 5.4 fb$^{\boldsymbol{-1}}$ of $\boldsymbol{p\bar{p}}$ collision
  data}}},\ \bibfield{journal}{%
  \Doi{10.1103/PhysRevD.83.032009}{\bibinfo {journal} {Phys. Rev. D}}\ }%
  \textbf{\bibinfo {volume} {83}},\ \bibinfo {pages} {032009} (\bibinfo {year}
  {2011}),\ \Eprint{http://arxiv.org/abs/1011.6549}{arXiv:1011.6549 [hep-ex]}.%
  \bibAnnoteFile{Stop}{Abazov:2010jn}%
\bibitem{Aad:2012ky}%
  \BibitemOpen
  \bibfield{author}{%
  \bibinfo {author} {\bibfnamefont{G.}~\bibnamefont{Aad}} \emph{et~al.}
  (\bibinfo {collaboration} {ATLAS Collaboration}),\ }%
  \emph{\bibinfo {title} {{Measurement of the W boson polarization in top quark
  decays with the ATLAS detector}}},\ \bibfield{journal}{%
  \Doi{10.1007/JHEP06(2012)088}{\bibinfo {journal} {JHEP}}\ }%
  \textbf{\bibinfo {volume} {1206}},\ \bibinfo {pages} {088} (\bibinfo {year}
  {2012}),\ \Eprint{http://arxiv.org/abs/1205.2484}{arXiv:1205.2484 [hep-ex]}.%
  \bibAnnoteFile{Stop}{Aad:2012ky}%
\bibitem{Chatrchyan:2013jna}%
  \BibitemOpen
  \bibfield{author}{%
  \bibinfo {author} {\bibfnamefont{S.}~\bibnamefont{Chatrchyan}} \emph{et~al.}
  (\bibinfo {collaboration} {CMS Collaboration}),\ }%
  \emph{\bibinfo {title} {{Measurement of the W-boson helicity in top-quark
  decays from ttbar production in lepton+jets events in pp collisions at
  sqrt(s)=7 TeV}}} (\bibinfo {year} {2013}),\
  \Eprint{http://arxiv.org/abs/1308.3879}{arXiv:1308.3879 [hep-ex]}.%
  \bibAnnoteFile{Stop}{Chatrchyan:2013jna}%
\bibitem{ATLAS-CONF-2013-032}%
  \BibitemOpen
  \enquote{\bibinfo {title} {{Search for $\cal{CP}$ violation in single top
  quark events in $pp$ collisions at $\sqrt{s} = 7$ TeV with the ATLAS
  detector}},}\ \bibinfo {howpublished} {ATLAS-CONF-2013-032},\ \bibinfo {note}
  {(2013)},\
  \url{http://cds.cern.ch/record/1527128/files/ATLAS-CONF-2013-032.}%
  \bibAnnoteFile{Stop}{ATLAS-CONF-2013-032}%
\bibitem{GonzalezSprinberg:2011kx}%
  \BibitemOpen
  \bibfield{author}{%
  \bibinfo {author} {\bibfnamefont{G.~A.}\ \bibnamefont{Gonz\'alez-Sprinberg}},
  \bibinfo {author} {\bibfnamefont{R.}~\bibnamefont{Mart{\'\i}nez}}\ y\
  \bibinfo {author} {\bibfnamefont{J.}~\bibnamefont{Vidal}},\ }%
  \emph{\bibinfo {title} {{Top quark tensor couplings}}},\ \bibfield{journal}{%
  \Doi{10.1007/JHEP07(2011)094}{\bibinfo {journal} {JHEP}}\ }%
  \textbf{\bibinfo {volume} {1107}},\ \bibinfo {pages} {094} (\bibinfo {year}
  {2011}),\ \Eprint{http://arxiv.org/abs/1105.5601}{arXiv:1105.5601 [hep-ph]}.%
  \bibAnnoteFile{Stop}{GonzalezSprinberg:2011kx}%
\bibitem{GonzalezSprinberg:2013er}%
  \BibitemOpen
  \bibfield{author}{%
  \bibinfo {author} {\bibfnamefont{G.~A.}\ \bibnamefont{Gonz\'alez-Sprinberg}},
  \bibinfo {author} {\bibfnamefont{R.}~\bibnamefont{Mart{\'\i}nez}}\ y\
  \bibinfo {author} {\bibfnamefont{J.}~\bibnamefont{Vidal}},\ }%
  \emph{\bibinfo {title} {{Erratum: Top quark tensor couplings}}},\
  \bibfield{journal}{%
  \Doi{10.1007/JHEP05(2013)117}{\bibinfo {journal} {JHEP}}\ }%
  \textbf{\bibinfo {volume} {1305}},\ \bibinfo {pages} {117} (\bibinfo {year}
  {2013}).%
  \bibAnnoteFile{Stop}{GonzalezSprinberg:2013er}%
\bibitem{Drobnak:2011aa}%
  \BibitemOpen
  \bibfield{author}{%
  \bibinfo {author} {\bibfnamefont{J.}~\bibnamefont{Drobnak}}, \bibinfo
  {author} {\bibfnamefont{S.}~\bibnamefont{Fajfer}}\ y\ \bibinfo {author}
  {\bibfnamefont{J.~F.}\ \bibnamefont{Kamenik}},\ }%
  \emph{\bibinfo {title} {{Probing anomalous tWb interactions with rare B
  decays}}},\ \bibfield{journal}{%
  \Doi{10.1016/j.nuclphysb.2011.10.004}{\bibinfo {journal} {Nucl. Phys. B}}\ }%
  \textbf{\bibinfo {volume} {855}},\ \bibinfo {pages} {82} (\bibinfo {year}
  {2012}),\ \Eprint{http://arxiv.org/abs/1109.2357}{arXiv:1109.2357 [hep-ph]}.%
  \bibAnnoteFile{Stop}{Drobnak:2011aa}%
\bibitem{Grzadkowski:2008mf}%
  \BibitemOpen
  \bibfield{author}{%
  \bibinfo {author} {\bibfnamefont{B.}~\bibnamefont{Grzadkowski}}\ y\ \bibinfo
  {author} {\bibfnamefont{M.}~\bibnamefont{Misiak}},\ }%
  \emph{\bibinfo {title} {{Anomalous Wtb coupling effects in the weak radiative
  B-meson decay}}},\ \bibfield{journal}{%
  \Doi{10.1103/PhysRevD.84.059903, 10.1103/PhysRevD.78.077501}{\bibinfo
  {journal} {Phys. Rev. D}}\ }%
  \textbf{\bibinfo {volume} {78}},\ \bibinfo {pages} {077501} (\bibinfo {year}
  {2008}),\ \Eprint{http://arxiv.org/abs/0802.1413}{arXiv:0802.1413 [hep-ph]}.%
  \bibAnnoteFile{Stop}{Grzadkowski:2008mf}%
\bibitem{AguilarSaavedra:2010nx}%
  \BibitemOpen
  \bibfield{author}{%
  \bibinfo {author} {\bibfnamefont{J.}~\bibnamefont{Aguilar-Saavedra}}\ y\
  \bibinfo {author} {\bibfnamefont{J.}~\bibnamefont{Bernab\'eu}},\ }%
  \emph{\bibinfo {title} {{W polarisation beyond helicity fractions in top
  quark decays}}},\ \bibfield{journal}{%
  \Doi{10.1016/j.nuclphysb.2010.07.012}{\bibinfo {journal} {Nucl. Phys. B}}\ }%
  \textbf{\bibinfo {volume} {840}},\ \bibinfo {pages} {349} (\bibinfo {year}
  {2010}),\ \Eprint{http://arxiv.org/abs/1005.5382}{arXiv:1005.5382 [hep-ph]}.%
  \bibAnnoteFile{Stop}{AguilarSaavedra:2010nx}%
\bibitem{Branco:2011iw}%
  \BibitemOpen
  \bibfield{author}{%
  \bibinfo {author} {\bibfnamefont{G.}~\bibnamefont{Branco}}, \bibinfo {author}
  {\bibfnamefont{P.}~\bibnamefont{Ferreira}}, \bibinfo {author}
  {\bibfnamefont{L.}~\bibnamefont{Lavoura}}, \bibinfo {author}
  {\bibfnamefont{M.}~\bibnamefont{Rebelo}}, \bibinfo {author}
  {\bibfnamefont{M.}~\bibnamefont{Sher}} \emph{et~al.},\ }%
  \emph{\bibinfo {title} {{Theory and phenomenology of two-Higgs-doublet
  models}}},\ \bibfield{journal}{%
  \Doi{10.1016/j.physrep.2012.02.002}{\bibinfo {journal} {Phys. Rept.}}\ }%
  \textbf{\bibinfo {volume} {516}},\ \bibinfo {pages} {1} (\bibinfo {year}
  {2012}),\ \Eprint{http://arxiv.org/abs/1106.0034}{arXiv:1106.0034 [hep-ph]}.%
  \bibAnnoteFile{Stop}{Branco:2011iw}%
\bibitem{Pich:2009sp}%
  \BibitemOpen
  \bibfield{author}{%
  \bibinfo {author} {\bibfnamefont{A.}~\bibnamefont{Pich}}\ y\ \bibinfo
  {author} {\bibfnamefont{P.}~\bibnamefont{Tuzon}},\ }%
  \emph{\bibinfo {title} {{Yukawa Alignment in the Two-Higgs-Doublet Model}}},\
  \bibfield{journal}{%
  \Doi{10.1103/PhysRevD.80.091702}{\bibinfo {journal} {Phys. Rev. D}}\ }%
  \textbf{\bibinfo {volume} {80}},\ \bibinfo {pages} {091702} (\bibinfo {year}
  {2009}),\ \Eprint{http://arxiv.org/abs/0908.1554}{arXiv:0908.1554 [hep-ph]}.%
  \bibAnnoteFile{Stop}{Pich:2009sp}%
\bibitem{Beneke:2000hk}%
  \BibitemOpen
  \bibfield{author}{%
  \bibinfo {author} {\bibfnamefont{M.}~\bibnamefont{Beneke}}, \bibinfo {author}
  {\bibfnamefont{I.}~\bibnamefont{Efthymiopoulos}}, \bibinfo {author}
  {\bibfnamefont{M.~L.}\ \bibnamefont{Mangano}}, \bibinfo {author}
  {\bibfnamefont{J.}~\bibnamefont{Womersley}}, \bibinfo {author}
  {\bibfnamefont{A.}~\bibnamefont{Ahmadov}} \emph{et~al.},\ }%
  \emph{\bibinfo {title} {{Top quark physics}}} (\bibinfo {year} {2000}),\
  \Eprint{http://arxiv.org/abs/hep-ph/0003033}{arXiv:hep-ph/0003033 [hep-ph]}.%
  \bibAnnoteFile{Stop}{Beneke:2000hk}%
\bibitem{Tait:2000sh}%
  \BibitemOpen
  \bibfield{author}{%
  \bibinfo {author} {\bibfnamefont{T.~M.}\ \bibnamefont{Tait}}\ y\ \bibinfo
  {author} {\bibfnamefont{C.-P.}\ \bibnamefont{Yuan}},\ }%
  \emph{\bibinfo {title} {{Single top quark production as a window to physics
  beyond the standard model}}},\ \bibfield{journal}{%
  \Doi{10.1103/PhysRevD.63.014018}{\bibinfo {journal} {Phys.Rev.}}\ }%
  \textbf{\bibinfo {volume} {D63}},\ \bibinfo {pages} {014018} (\bibinfo {year}
  {2000}),\ \Eprint{http://arxiv.org/abs/hep-ph/0007298}{arXiv:hep-ph/0007298
  [hep-ph]}.%
  \bibAnnoteFile{Stop}{Tait:2000sh}%
\bibitem{Bernreuther:2008us}%
  \BibitemOpen
  \bibfield{author}{%
  \bibinfo {author} {\bibfnamefont{W.}~\bibnamefont{Bernreuther}}, \bibinfo
  {author} {\bibfnamefont{P.}~\bibnamefont{Gonzalez}}\ y\ \bibinfo {author}
  {\bibfnamefont{M.}~\bibnamefont{Wiebusch}},\ }%
  \emph{\bibinfo {title} {{The Top Quark Decay Vertex in Standard Model
  Extensions}}},\ \bibfield{journal}{%
  \Doi{10.1140/epjc/s10052-009-0887-4}{\bibinfo {journal} {Eur. Phys. J. C}}\
  }%
  \textbf{\bibinfo {volume} {60}},\ \bibinfo {pages} {197} (\bibinfo {year}
  {2009}),\ \Eprint{http://arxiv.org/abs/0812.1643}{arXiv:0812.1643 [hep-ph]}.%
  \bibAnnoteFile{Stop}{Bernreuther:2008us}%
\bibitem{Jung:2010ik}%
  \BibitemOpen
  \bibfield{author}{%
  \bibinfo {author} {\bibfnamefont{M.}~\bibnamefont{Jung}}, \bibinfo {author}
  {\bibfnamefont{A.}~\bibnamefont{Pich}}\ y\ \bibinfo {author}
  {\bibfnamefont{P.}~\bibnamefont{Tuzon}},\ }%
  \emph{\bibinfo {title} {{Charged-Higgs phenomenology in the Aligned
  two-Higgs-doublet model}}},\ \bibfield{journal}{%
  \Doi{10.1007/JHEP11(2010)003}{\bibinfo {journal} {JHEP}}\ }%
  \textbf{\bibinfo {volume} {1011}},\ \bibinfo {pages} {003} (\bibinfo {year}
  {2010}),\ \Eprint{http://arxiv.org/abs/1006.0470}{arXiv:1006.0470 [hep-ph]}.%
  \bibAnnoteFile{Stop}{Jung:2010ik}%
\bibitem{Jung:2010ab}%
  \BibitemOpen
  \bibfield{author}{%
  \bibinfo {author} {\bibfnamefont{M.}~\bibnamefont{Jung}}, \bibinfo {author}
  {\bibfnamefont{A.}~\bibnamefont{Pich}}\ y\ \bibinfo {author}
  {\bibfnamefont{P.}~\bibnamefont{Tuzon}},\ }%
  \emph{\bibinfo {title} {{The $b \to X_{s} \gamma$ Rate and CP Asymmetry
  within the Aligned Two-Higgs-Doublet Model}}},\ \bibfield{journal}{%
  \Doi{10.1103/PhysRevD.83.074011}{\bibinfo {journal} {Phys. Rev. D}}\ }%
  \textbf{\bibinfo {volume} {83}},\ \bibinfo {pages} {074011} (\bibinfo {year}
  {2011}),\ \Eprint{http://arxiv.org/abs/1011.5154}{arXiv:1011.5154 [hep-ph]}.%
  \bibAnnoteFile{Stop}{Jung:2010ab}%
\bibitem{Jung:2012vu}%
  \BibitemOpen
  \bibfield{author}{%
  \bibinfo {author} {\bibfnamefont{M.}~\bibnamefont{Jung}}, \bibinfo {author}
  {\bibfnamefont{X.-Q.}\ \bibnamefont{Li}}\ y\ \bibinfo {author}
  {\bibfnamefont{A.}~\bibnamefont{Pich}},\ }%
  \emph{\bibinfo {title} {{Exclusive radiative B-meson decays within the
  aligned two-Higgs-doublet model}}},\ \bibfield{journal}{%
  \Doi{10.1007/JHEP10(2012)063}{\bibinfo {journal} {JHEP}}\ }%
  \textbf{\bibinfo {volume} {1210}},\ \bibinfo {pages} {063} (\bibinfo {year}
  {2012}),\ \Eprint{http://arxiv.org/abs/1208.1251}{arXiv:1208.1251 [hep-ph]}.%
  \bibAnnoteFile{Stop}{Jung:2012vu}%
\bibitem{Celis:2012dk}%
  \BibitemOpen
  \bibfield{author}{%
  \bibinfo {author} {\bibfnamefont{A.}~\bibnamefont{Celis}}, \bibinfo {author}
  {\bibfnamefont{M.}~\bibnamefont{Jung}}, \bibinfo {author}
  {\bibfnamefont{X.-Q.}\ \bibnamefont{Li}}\ y\ \bibinfo {author}
  {\bibfnamefont{A.}~\bibnamefont{Pich}},\ }%
  \emph{\bibinfo {title} {{Sensitivity to charged scalars in $\boldsymbol{B\to
  D^{(*)}\tau\nu_\tau}$ and $\boldsymbol{B\to\tau\nu_\tau}$ decays}}},\
  \bibfield{journal}{%
  \Doi{10.1007/JHEP01(2013)054}{\bibinfo {journal} {JHEP}}\ }%
  \textbf{\bibinfo {volume} {1301}},\ \bibinfo {pages} {054} (\bibinfo {year}
  {2013}),\ \Eprint{http://arxiv.org/abs/1210.8443}{arXiv:1210.8443 [hep-ph]}.%
  \bibAnnoteFile{Stop}{Celis:2012dk}%
\bibitem{Celis:2013rcs}%
  \BibitemOpen
  \bibfield{author}{%
  \bibinfo {author} {\bibfnamefont{A.}~\bibnamefont{Celis}}, \bibinfo {author}
  {\bibfnamefont{V.}~\bibnamefont{Ilisie}}\ y\ \bibinfo {author}
  {\bibfnamefont{A.}~\bibnamefont{Pich}},\ }%
  \emph{\bibinfo {title} {{LHC constraints on two-Higgs doublet models}}},\
  \bibfield{journal}{%
  \Doi{10.1007/JHEP07(2013)053}{\bibinfo {journal} {JHEP}}\ }%
  \textbf{\bibinfo {volume} {1307}},\ \bibinfo {pages} {053} (\bibinfo {year}
  {2013}),\ \Eprint{http://arxiv.org/abs/1302.4022}{arXiv:1302.4022 [hep-ph]}.%
  \bibAnnoteFile{Stop}{Celis:2013rcs}%
\bibitem{Lees:2012xj}%
  \BibitemOpen
  \bibfield{author}{%
  \bibinfo {author} {\bibfnamefont{J.}~\bibnamefont{Lees}} \emph{et~al.}
  (\bibinfo {collaboration} {BaBar Collaboration}),\ }%
  \emph{\bibinfo {title} {{Evidence for an excess of $\bar{B} \to D^{(*)}
  \tau^-\bar{\nu}_\tau$ decays}}},\ \bibfield{journal}{%
  \Doi{10.1103/PhysRevLett.109.101802}{\bibinfo {journal} {Phys.Rev.Lett.}}\ }%
  \textbf{\bibinfo {volume} {109}},\ \bibinfo {pages} {101802} (\bibinfo {year}
  {2012}),\ \Eprint{http://arxiv.org/abs/1205.5442}{arXiv:1205.5442 [hep-ex]}.%
  \bibAnnoteFile{Stop}{Lees:2012xj}%
\bibitem{Aaltonen:2012qt}%
  \BibitemOpen
  \bibfield{author}{%
  \bibinfo {author} {\bibfnamefont{T.}~\bibnamefont{Aaltonen}} \emph{et~al.}
  (\bibinfo {collaboration} {CDF Collaboration, D0 Collaboration}),\ }%
  \emph{\bibinfo {title} {{Evidence for a particle produced in association with
  weak bosons and decaying to a bottom-antibottom quark pair in Higgs boson
  searches at the Tevatron}}},\ \bibfield{journal}{%
  \Doi{10.1103/PhysRevLett.109.071804}{\bibinfo {journal} {Phys.Rev.Lett.}}\ }%
  \textbf{\bibinfo {volume} {109}},\ \bibinfo {pages} {071804} (\bibinfo {year}
  {2012}),\ \Eprint{http://arxiv.org/abs/1207.6436}{arXiv:1207.6436 [hep-ex]}.%
  \bibAnnoteFile{Stop}{Aaltonen:2012qt}%
\bibitem{Buchmuller:1985jz}%
  \BibitemOpen
  \bibfield{author}{%
  \bibinfo {author} {\bibfnamefont{W.}~\bibnamefont{Buchmuller}}\ y\ \bibinfo
  {author} {\bibfnamefont{D.}~\bibnamefont{Wyler}},\ }%
  \emph{\bibinfo {title} {{Effective Lagrangian Analysis of New Interactions
  and Flavor Conservation}}},\ \bibfield{journal}{%
  \Doi{10.1016/0550-3213(86)90262-2}{\bibinfo {journal} {Nucl. Phys. B}}\ }%
  \textbf{\bibinfo {volume} {268}},\ \bibinfo {pages} {621} (\bibinfo {year}
  {1986}).%
  \bibAnnoteFile{Stop}{Buchmuller:1985jz}%
\bibitem{AguilarSaavedra:2008zc}%
  \BibitemOpen
  \bibfield{author}{%
  \bibinfo {author} {\bibfnamefont{J.}~\bibnamefont{Aguilar-Saavedra}},\ }%
  \emph{\bibinfo {title} {{A Minimal set of top anomalous couplings}}},\
  \bibfield{journal}{%
  \Doi{10.1016/j.nuclphysb.2008.12.012}{\bibinfo {journal} {Nucl. Phys. B}}\ }%
  \textbf{\bibinfo {volume} {812}},\ \bibinfo {pages} {181} (\bibinfo {year}
  {2009}),\ \Eprint{http://arxiv.org/abs/0811.3842}{arXiv:0811.3842 [hep-ph]}.%
  \bibAnnoteFile{Stop}{AguilarSaavedra:2008zc}%
\bibitem{Kane:1991bg}%
  \BibitemOpen
  \bibfield{author}{%
  \bibinfo {author} {\bibfnamefont{G.~L.}\ \bibnamefont{Kane}}, \bibinfo
  {author} {\bibfnamefont{G.}~\bibnamefont{Ladinsky}}\ y\ \bibinfo {author}
  {\bibfnamefont{C.}~\bibnamefont{Yuan}},\ }%
  \emph{\bibinfo {title} {{Using the Top Quark for Testing Standard Model
  Polarization and CP Predictions}}},\ \bibfield{journal}{%
  \Doi{10.1103/PhysRevD.45.124}{\bibinfo {journal} {Phys. Rev. D}}\ }%
  \textbf{\bibinfo {volume} {45}},\ \bibinfo {pages} {124} (\bibinfo {year}
  {1992}).%
  \bibAnnoteFile{Stop}{Kane:1991bg}%
\bibitem{Beringer:1900zz}%
  \BibitemOpen
  \bibfield{author}{%
  \bibinfo {author} {\bibfnamefont{J.}~\bibnamefont{Beringer}} \emph{et~al.}
  (\bibinfo {collaboration} {Particle Data Group}),\ }%
  \emph{\bibinfo {title} {{Review of Particle Physics (RPP)}}},\
  \bibfield{journal}{%
  \Doi{10.1103/PhysRevD.86.010001}{\bibinfo {journal} {Phys. Rev. D}}\ }%
  \textbf{\bibinfo {volume} {86}},\ \bibinfo {pages} {010001} (\bibinfo {year}
  {2012}).%
  \bibAnnoteFile{Stop}{Beringer:1900zz}%
\bibitem{Li:1990qf}%
  \BibitemOpen
  \bibfield{author}{%
  \bibinfo {author} {\bibfnamefont{C.~S.}\ \bibnamefont{Li}}, \bibinfo {author}
  {\bibfnamefont{R.~J.}\ \bibnamefont{Oakes}}\ y\ \bibinfo {author}
  {\bibfnamefont{T.~C.}\ \bibnamefont{Yuan}},\ }%
  \emph{\bibinfo {title} {{QCD corrections to $t \to W^{+} b$}}},\
  \bibfield{journal}{%
  \Doi{10.1103/PhysRevD.43.3759}{\bibinfo {journal} {Phys. Rev. D}}\ }%
  \textbf{\bibinfo {volume} {43}},\ \bibinfo {pages} {3759} (\bibinfo {year}
  {1991}).%
  \bibAnnoteFile{Stop}{Li:1990qf}%
\bibitem{Lampe:1995xb}%
  \BibitemOpen
  \bibfield{author}{%
  \bibinfo {author} {\bibfnamefont{B.}~\bibnamefont{Lampe}},\ }%
  \emph{\bibinfo {title} {{Forward - backward asymmetry in top quark
  semileptonic decay}}},\ \bibfield{journal}{%
  \Doi{10.1016/0550-3213(95)00420-W}{\bibinfo {journal} {Nucl.Phys.}}\ }%
  \textbf{\bibinfo {volume} {B454}},\ \bibinfo {pages} {506} (\bibinfo {year}
  {1995}).%
  \bibAnnoteFile{Stop}{Lampe:1995xb}%
\bibitem{delAguila:2002nf}%
  \BibitemOpen
  \bibfield{author}{%
  \bibinfo {author} {\bibfnamefont{F.}~\bibnamefont{del Aguila}}\ y\ \bibinfo
  {author} {\bibfnamefont{J.}~\bibnamefont{Aguilar-Saavedra}},\ }%
  \emph{\bibinfo {title} {{Precise determination of the Wtb couplings at CERN
  LHC}}},\ \bibfield{journal}{%
  \Doi{10.1103/PhysRevD.67.014009}{\bibinfo {journal} {Phys.Rev.}}\ }%
  \textbf{\bibinfo {volume} {D67}},\ \bibinfo {pages} {014009} (\bibinfo {year}
  {2003}),\ \Eprint{http://arxiv.org/abs/hep-ph/0208171}{arXiv:hep-ph/0208171
  [hep-ph]}.%
  \bibAnnoteFile{Stop}{delAguila:2002nf}%
\bibitem{Rindani:2011pk}%
  \BibitemOpen
  \bibfield{author}{%
  \bibinfo {author} {\bibfnamefont{S.~D.}\ \bibnamefont{Rindani}}\ y\ \bibinfo
  {author} {\bibfnamefont{P.}~\bibnamefont{Sharma}},\ }%
  \emph{\bibinfo {title} {{Probing anomalous tbW couplings in single-top
  production using top polarization at the Large Hadron Collider}}},\
  \bibfield{journal}{%
  \Doi{10.1007/JHEP11(2011)082}{\bibinfo {journal} {JHEP}}\ }%
  \textbf{\bibinfo {volume} {1111}},\ \bibinfo {pages} {082} (\bibinfo {year}
  {2011}),\ \Eprint{http://arxiv.org/abs/1107.2597}{arXiv:1107.2597 [hep-ph]}.%
  \bibAnnoteFile{Stop}{Rindani:2011pk}%
\bibitem{:2012iwa}%
  \BibitemOpen
  \bibfield{author}{%
  \bibinfo {author} {\bibfnamefont{V.~M.}\ \bibnamefont{Abazov}} \emph{et~al.}
  (\bibinfo {collaboration} {D0 Collaboration}),\ }%
  \emph{\bibinfo {title} {{Combination of searches for anomalous top quark
  couplings with 5.4 fb$^{-1}$ of $p\bar{p}$ collisions}}},\
  \bibfield{journal}{%
  \Doi{10.1016/j.physletb.2012.05.048}{\bibinfo {journal} {Phys. Lett. B}}\ }%
  \textbf{\bibinfo {volume} {713}},\ \bibinfo {pages} {165} (\bibinfo {year}
  {2012}),\ \Eprint{http://arxiv.org/abs/1204.2332}{arXiv:1204.2332 [hep-ex]}.%
  \bibAnnoteFile{Stop}{:2012iwa}%
\bibitem{Abazov:2011pm}%
  \BibitemOpen
  \bibfield{author}{%
  \bibinfo {author} {\bibfnamefont{V.~M.}\ \bibnamefont{Abazov}} \emph{et~al.}
  (\bibinfo {collaboration} {D0 Collaboration}),\ }%
  \emph{\bibinfo {title} {{Search for anomalous $Wtb$ couplings in single top
  quark production in $p\bar{p}$ collisions at $\sqrt{s} = 1.96$ TeV}}},\
  \bibfield{journal}{%
  \Doi{10.1016/j.physletb.2012.01.014}{\bibinfo {journal} {Phys. Lett. B}}\ }%
  \textbf{\bibinfo {volume} {708}},\ \bibinfo {pages} {21} (\bibinfo {year}
  {2012}),\ \Eprint{http://arxiv.org/abs/1110.4592}{arXiv:1110.4592 [hep-ex]}.%
  \bibAnnoteFile{Stop}{Abazov:2011pm}%
\bibitem{AguilarSaavedra:2011ct}%
  \BibitemOpen
  \bibfield{author}{%
  \bibinfo {author} {\bibfnamefont{J.}~\bibnamefont{Aguilar-Saavedra}},
  \bibinfo {author} {\bibfnamefont{N.}~\bibnamefont{Castro}}\ y\ \bibinfo
  {author} {\bibfnamefont{A.}~\bibnamefont{Onofre}},\ }%
  \emph{\bibinfo {title} {{Constraints on the Wtb vertex from early LHC
  data}}},\ \bibfield{journal}{%
  \Doi{10.1103/PhysRevD.84.019901, 10.1103/PhysRevD.83.117301}{\bibinfo
  {journal} {Phys. Rev. D}}\ }%
  \textbf{\bibinfo {volume} {83}},\ \bibinfo {pages} {117301} (\bibinfo {year}
  {2011}),\ \Eprint{http://arxiv.org/abs/1105.0117}{arXiv:1105.0117 [hep-ph]}.%
  \bibAnnoteFile{Stop}{AguilarSaavedra:2011ct}%
\bibitem{Ligeti:2010ia}%
  \BibitemOpen
  \bibfield{author}{%
  \bibinfo {author} {\bibfnamefont{Z.}~\bibnamefont{Ligeti}}, \bibinfo {author}
  {\bibfnamefont{M.}~\bibnamefont{Papucci}}, \bibinfo {author}
  {\bibfnamefont{G.}~\bibnamefont{Perez}}\ y\ \bibinfo {author}
  {\bibfnamefont{J.}~\bibnamefont{Zupan}},\ }%
  \emph{\bibinfo {title} {{Implication s of the dimuon CP asymmetry in
  $B_{d,s}$ decays}}},\ \bibfield{journal}{%
  \Doi{10.1103/PhysRevLett.105.131601}{\bibinfo {journal} {Phys. Rev. Lett.}}\
  }%
  \textbf{\bibinfo {volume} {105}},\ \bibinfo {pages} {131601} (\bibinfo {year}
  {2010}),\ \Eprint{http://arxiv.org/abs/1006.0432}{arXiv:1006.0432 [hep-ph]}.%
  \bibAnnoteFile{Stop}{Ligeti:2010ia}%
\bibitem{Lenz:2010gu}%
  \BibitemOpen
  \bibfield{author}{%
  \bibinfo {author} {\bibfnamefont{A.}~\bibnamefont{Lenz}}, \bibinfo {author}
  {\bibfnamefont{U.}~\bibnamefont{Nierste}}, \bibinfo {author}
  {\bibfnamefont{J.}~\bibnamefont{Charles}}, \bibinfo {author}
  {\bibfnamefont{S.}~\bibnamefont{Descotes-Genon}}, \bibinfo {author}
  {\bibfnamefont{A.}~\bibnamefont{Jantsch}} \emph{et~al.},\ }%
  \emph{\bibinfo {title} {{Anatomy of New Physics in $B - \bar{B}$ mixing}}},\
  \bibfield{journal}{%
  \Doi{10.1103/PhysRevD.83.036004}{\bibinfo {journal} {Phys. Rev. D}}\ }%
  \textbf{\bibinfo {volume} {83}},\ \bibinfo {pages} {036004} (\bibinfo {year}
  {2011}),\ \Eprint{http://arxiv.org/abs/1008.1593}{arXiv:1008.1593 [hep-ph]}.%
  \bibAnnoteFile{Stop}{Lenz:2010gu}%
\bibitem{Aaltonen:2011rj}%
  \BibitemOpen
  \bibfield{author}{%
  \bibinfo {author} {\bibfnamefont{T.}~\bibnamefont{Aaltonen}} \emph{et~al.}
  (\bibinfo {collaboration} {CDF Collaboration}),\ }%
  \emph{\bibinfo {title} {{Search for a Higgs Boson in the Diphoton Final State
  in $p\bar{p}$ Collisions at $\sqrt{s} = 1.96$ TeV}}},\ \bibfield{journal}{%
  \Doi{10.1103/PhysRevLett.108.011801}{\bibinfo {journal} {Phys. Rev. Lett.}}\
  }%
  \textbf{\bibinfo {volume} {108}},\ \bibinfo {pages} {011801} (\bibinfo {year}
  {2012}),\ \Eprint{http://arxiv.org/abs/1109.4427}{arXiv:1109.4427 [hep-ex]}.%
  \bibAnnoteFile{Stop}{Aaltonen:2011rj}%
\bibitem{Abazov:2011ix}%
  \BibitemOpen
  \bibfield{author}{%
  \bibinfo {author} {\bibfnamefont{V.}~\bibnamefont{Abazov}} \emph{et~al.}
  (\bibinfo {collaboration} {D0 Collaboration}),\ }%
  \emph{\bibinfo {title} {{Search for the standard model and a fermiophobic
  Higgs boson in diphoton final states}}},\ \bibfield{journal}{%
  \Doi{10.1103/PhysRevLett.107.151801}{\bibinfo {journal} {Phys. Rev. Lett.}}\
  }%
  \textbf{\bibinfo {volume} {107}},\ \bibinfo {pages} {151801} (\bibinfo {year}
  {2011}),\ \Eprint{http://arxiv.org/abs/1107.4587}{arXiv:1107.4587 [hep-ex]}.%
  \bibAnnoteFile{Stop}{Abazov:2011ix}%
\bibitem{Gunion:1989we}%
  \BibitemOpen
  \bibfield{author}{%
  \bibinfo {author} {\bibfnamefont{J.~F.}\ \bibnamefont{Gunion}}, \bibinfo
  {author} {\bibfnamefont{H.~E.}\ \bibnamefont{Haber}}, \bibinfo {author}
  {\bibfnamefont{G.~L.}\ \bibnamefont{Kane}}\ y\ \bibinfo {author}
  {\bibfnamefont{S.}~\bibnamefont{Dawson}},\ }%
  \emph{\bibinfo {title} {{THE HIGGS HUNTER'S GUIDE}}},\ \bibfield{journal}{%
  \bibinfo {journal} {Front. Phys.}\ }%
  \textbf{\bibinfo {volume} {80}},\ \bibinfo {pages} {1} (\bibinfo {year}
  {2000}).%
  \bibAnnoteFile{Stop}{Gunion:1989we}%
\bibitem{AguilarSaavedra:2007rs}%
  \BibitemOpen
  \bibfield{author}{%
  \bibinfo {author} {\bibfnamefont{J.}~\bibnamefont{Aguilar-Saavedra}},
  \bibinfo {author} {\bibfnamefont{J.}~\bibnamefont{Carvalho}}, \bibinfo
  {author} {\bibfnamefont{N.~F.}\ \bibnamefont{Castro}}, \bibinfo {author}
  {\bibfnamefont{A.}~\bibnamefont{Onofre}}\ y\ \bibinfo {author}
  {\bibfnamefont{F.}~\bibnamefont{Veloso}},\ }%
  \emph{\bibinfo {title} {{ATLAS sensitivity to Wtb anomalous couplings in top
  quark decays}}},\ \bibfield{journal}{%
  \Doi{10.1140/epjc/s10052-007-0519-9}{\bibinfo {journal} {Eur.Phys.J.}}\ }%
  \textbf{\bibinfo {volume} {C53}},\ \bibinfo {pages} {689} (\bibinfo {year}
  {2008}),\ \Eprint{http://arxiv.org/abs/0705.3041}{arXiv:0705.3041 [hep-ph]}.%
  \bibAnnoteFile{Stop}{AguilarSaavedra:2007rs}%
\end{thebibliography}%

\end{document}